# Inferring protein folding mechanisms from natural sequence diversity

Ezequiel A. Galpern[1], Ernesto A. Roman[2] and Diego U. Ferreiro[1*]

[1] Laboratorio de Fisiología de Proteínas, Facultad de Ciencias Exactas y Naturales, Universidad de Buenos Aires and Consejo Nacional de Investigaciones Científicas y Técnicas, Instituto de Química Biológica de la Facultad de Ciencias Exactas y Naturales (IQUIBICEN-CONICET), Buenos Aires CP1428, Argentina.

[2] Laboratorio de Ingeniería Enzimática y Nanobiotecnología, Facultad de Ciencias Exactas y Naturales, Universidad de Buenos Aires and Consejo Nacional de Investigaciones Científicas y Técnicas, Instituto de Química Biológica de la Facultad de Ciencias Exactas y Naturales (IQUIBICEN-CONICET), Buenos Aires CP1428, Argentina.

* To whom correspondence may be addressed: ferreiro@qb.fcen.uba.ar

Running Title: Sequence variations and protein folding

## Abstract

Protein sequences serve as a natural record of the evolutionary constraints that shape their functional structures. We show that it is possible to use only sequence information to go beyond predicting native structures and global stability to infer the folding mechanisms of globular proteins. The one- and two-body evolutionary energy fields at the amino-acid level are mapped to a coarse-grained description of folding, where proteins are divided into contiguous folding elements, commonly referred to as foldons. For 15 diverse protein families, we calculated the folding mechanisms of hundreds of proteins by simulating an Ising chain of foldons, with their energetics determined by the amino acid sequences. We show that protein topology imposes limits on the variability of folding cooperativity within a family. While most beta and alpha/beta structures exhibit only a few possible mechanisms despite high sequence diversity, alpha topologies allow for diverse folding scenarios among family members. We show that both the stability and cooperativity changes induced by mutations can be computed directly using sequence-based evolutionary models.

## Statement of significance

Closely related proteins usually share similar three-dimensional structures. Differences in their amino acid sequences can lead to distinct folding mechanisms, enabling the natural evolution of diverse biological functions. In this study, we developed a simplified model of protein folding by dividing proteins into folding elements. By using only sequence data, we simulated the folding of thousands of proteins. We found that the structural topology shared within a protein family determines whether diverse folding mechanisms can arise along the family members and predict the effect of mutations.

## Introduction

The original mysteries of protein folding are now well-understood. Energy Landscape Theory provides a deep theoretical understanding of how protein folding happens and how experiments can be interpreted [1]. The fundamental insight was the recognition that natural protein molecules are minimally frustrated heteropolymers, and therefore the overall shape of their energy landscape is that of a rough funnel [2]. This distinguishes them from most random, unfoldable, heteropolymers and thus the fact that natural protein molecules fold rapidly and robustly must be the outcome of evolution [3]. Conversely, protein sequences are believed to evolve in landscapes that cannot be excessively rugged, as this would hinder efficient exploration of novel structures [4]. In addition, the particular shape of the physical folding landscape constrains the natural exploration of the sequence space [5]. Today we recognize that natural protein molecules are foldable and evolvable systems whose functional structures can be *coded* in linear strings of monomers [6]. A precise description of how this *coding* is achieved is still unclear [7]. However, throughout their natural history, proteins have suffered countless number of random mutation events where folding landscapes have been *de facto* explored [8]. Studying the patterns emerging from comparative sequence analysis have been one of the keys -for humans and machines- to capture and learn at least some aspects of the folding *codes* [9]. Here we will use evolutionary data to inform a folding model and explore how topology and sequence variations impact on the folding mechanism of various protein families.

Folding of most single-domain proteins can be reasonably well-predicted with purely topological models [10], [11]. Although not perfect, the appearance of folding intermediates, the structures of transition states and the folding speeds are usually recapitulated with structure-based models that take a native backbone structure as sole input. Variations in the energetics of contact-based potentials have been shown to refine the correspondence with experiments, but this was necessarily done on anecdotal cases [12]. Alternatively, toy models such as lattice proteins have been used for assessing sequence impact on folding mechanisms and their relationship with evolution [13].

Only recently the sheer amount of sequence data allowed for the inference of amino acid interaction potentials based on maximum entropy Potts-like models, pioneered by Direct Coupling Analysis (DCA) [14], [15]. These representations make use of observed evolutionary sequence variations in a protein family to model the pairwise energetic coupling between positions [16]. The derived one-body and two-body terms can be interpreted as an evolutionary energetic field and can be used to infer native contacts [17]. Moreover, this ‘evolutionary energy’ can be evaluated for any sequence variation within a protein family and it has been shown to correlate with their folding energy variations [18], [19]. Also, the study of the fitness landscapes inferred by DCA models have provided insights on the evolution of protein sequences [20], [21], [22], [23]. It is generally believed that 'biological function' is the main evolutionary driving force acting on protein sequences, yet this includes many diverse activities such as catalysis, binding, allostery, avoidance of aggregation and so forth, which can conflict with robust folding of protein domains [6]. Here we will make the approximation that the energetics of protein folding is the main evolutionary pressure acting globally on protein sequences, which is in line with the finding that the average information contained in the sequences equals the average information needed to specify a fold [7].

When analyzing the folding of single-domain proteins, it was recognized that different protein regions may fold at different times quasi-independently if there are sufficiently strong native interactions within them to overcome their entropy loss. These units then may fold in a single cooperative step and have been christened *foldons* by Panchenko et al. [24]. It has been recently shown that conserved exons manifest a pronounced independent foldability, supporting the exon-foldon correspondence [25]. Leveraging on the annotation of the intron-exon boundaries in several genomes, we propose here to use this partitioning of the primary structure to define common foldons for each family and analyze their folding dynamic.

Mapping an inferred evolutionary energy to a folding model was previously done for repeat-proteins [26]. In these proteins the structural symmetry allows a natural way to define folding units. By modeling the interactions between these minimal common foldons, different groups of elements that fold at the same time emerge naturally for each protein, defining domains that coincide with those described experimentally. It was found that natural sequence variations enable a diversity of folding mechanisms, allowed by the elongated topology of the systems. Here we will examine to which extent the local energetic differences given by sequence modifications perturb the global aspects of the folding mechanism in globular domains.

## Results and Discussion

### *Model Definition*

We propose a generalization of a coarse-grained folding model that we have previously introduced for repeat-proteins [26], [27] to aperiodic topologies. We consider a protein as an array of interacting folding elements (foldons) that can be either folded (F) or unfolded (U), as 2-state spin variables. The system is represented as a finite-size Ising chain of $N$ elements, where the energy of a coarse-grained configuration, the Hamiltonian, is given by the free energy of the corresponding ensemble of microstates

$$H = -\sum_{j=1}^{N}[T s_j(1 - \delta_{j,F}) + \epsilon_j^i \delta_{j,F}] - \sum_{j=1}^{N-1}\sum_{k>j} \epsilon_{jk}^s \delta_{j,F}\delta_{k,F}, \quad (1)$$

where $T$ is the temperature and $\delta_{j,F}$ is the Kroeneker symbol taking value one if element $j$ is folded (F) and zero otherwise. If the element $j$ is folded, it has a specific internal folding free energy (averaged over the solvent) $\epsilon_j^i$. If two elements $j$ and $k$ are both folded, we consider also a surface energy $\epsilon_{jk}^s$, describing a specific interaction between the two foldons. If the element $j$ is unfolded we set the energetic contributions to zero, but there is an explicit entropic contribution given by the entropy $s_j$ of the available spatial configurations of the foldon. Hence, within this model a protein can unfold as a result of an increase in temperature $T$.

To apply this model effectively, two key tasks are essential. The first involves dividing the protein sequence into distinct, non-overlapping elements, or foldons, which are groups of amino acids that fold and unfold collectively. In this work, we leveraged the natural division provided by exon-intron structures, a natural division of genes into pieces (Fig. 1). For Multiple Sequence Alignments (MSA) of diverse protein families, we have found that the positions of the exon boundaries are highly conserved, allowing a consistent partition of each MSA into Minimal Common Exons (MCE) that we use here as protein folding elements. For some families, MCEs also match secondary structure elements. In addition, we have demonstrated that there is a pronounced tendency to independent foldability for protein segments corresponding to the more conserved exons, supporting an exon–foldon correspondence [25].

The second task is to assign energetic ($\epsilon_j^i$, $\epsilon_{jk}^s$) and the entropic ($s_j$) parameters as functions of the amino acid sequence $\boldsymbol{\sigma}$. We used a sequence-based evolutionary energy field to calculate the folding energy terms [26]. In particular, we learned a Restricted Boltzmann Machine (RBM) model [28] for each protein family. We map the obtained parameters to a Potts model, obtaining the residue–residue couplings $J_{ab}(\sigma_a, \sigma_b)$ and local fields $h_a(\sigma_a)$ to calculate the specific coarse-grained folding free energy terms for each sequence,

$$\epsilon_j^i = \epsilon^i(\boldsymbol{\sigma}_j) = k_B T_{sel}\left[\sum_{a \in \mathcal{A}_j} h_a(\sigma_a) + \sum_{a,b \in \mathcal{A}_j} J_{ab}(\sigma_a, \sigma_b)\right], \quad (2a)$$

$$\epsilon_{kj}^s = \epsilon^s(\boldsymbol{\sigma}_j, \boldsymbol{\sigma}_k) = k_B T_{sel}\left[\sum_{a \in \mathcal{A}_j,\, b \in \mathcal{A}_k} J_{ab}(\sigma_a, \sigma_b)\right], \quad (2b)$$

where $\mathcal{A}_j$ is the set of amino acid positions in foldon $j$, $k_B$ is the Boltzmann constant and $T_{sel}$ is the Selection Temperature. $T_{sel}$ is the apparent temperature at which sequences of a particular family were selected by nature and quantifies how strong the folding constraints have been during evolution [29]. Given a family, $k_B T_{sel}$ can be explicitly calculated as the proportionality constant between the folding free energy changes upon mutations $\Delta\Delta G$ and the corresponding changes in the evolutionary energy. In this work, we used experimental $\Delta\Delta G$ for calculating $T_{sel}^{PDZ}$ for a reference, PDZ family, and leveraged Miyazawa's observation [30] to estimate $T_{sel}$ for other families. Assuming that the standard deviation of $\Delta\Delta G$ is nearly constant irrespectively of protein families, $T_{sel}$ is scaled relative to $T_{sel}^{PDZ}$, using the ratio of the standard deviation of evolutionary energy changes by single mutations.  To compute the entropic terms, we take $s_j$ to be independent of amino acid identity and strictly additive and we use the average entropy per residue fitted for the repeat-protein model [26]. Therefore $s_j = L_j s$, being $L_j$ the sequence length of the element and $s = 5\, cal\, mol^{-1} K^{-1} res^{-1}$.

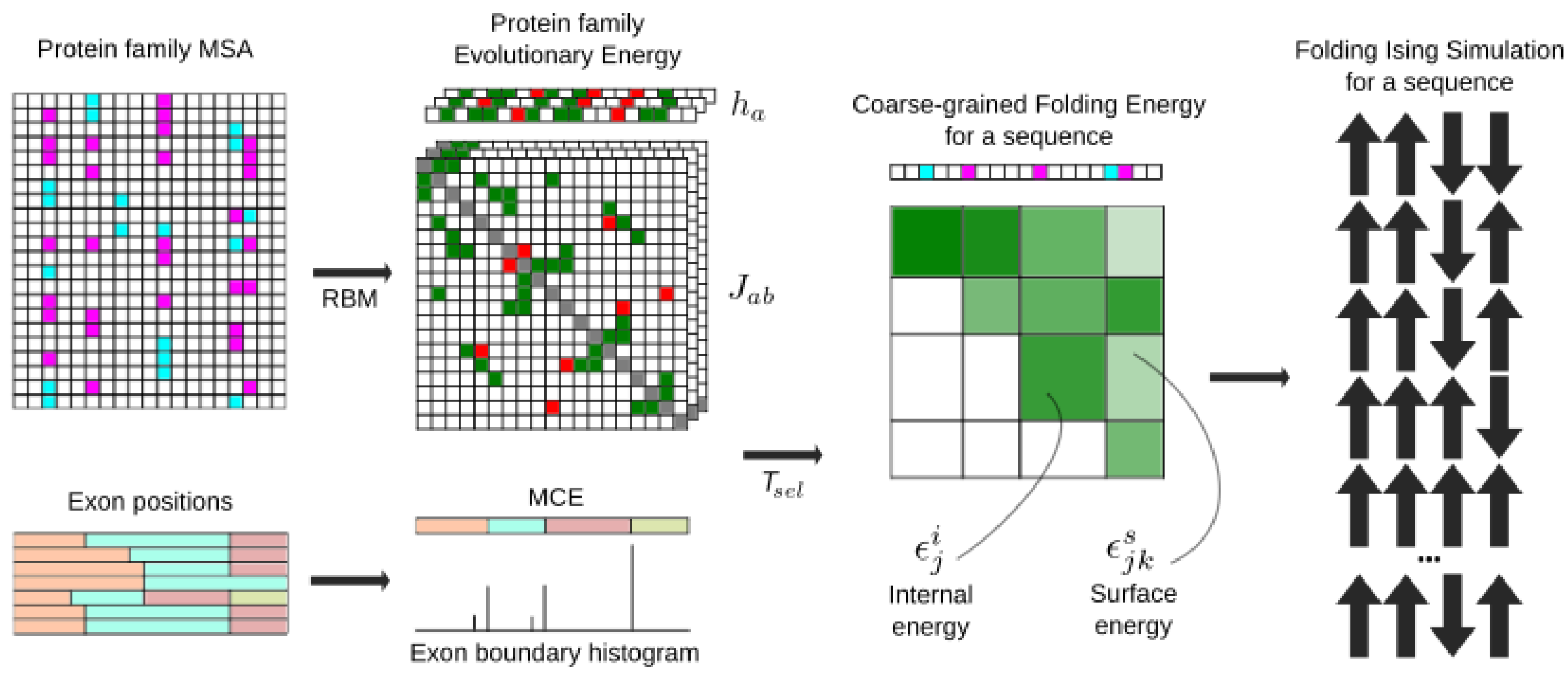

***Figure 1. Model Definition.*** *We learn the Evolutionary Energy field parameters from a Multiple Sequence Alignment (MSA). Using as folding units the Minimal Common Exons (MCE) [25] and the Selection Temperature of the family, we extract the coarse-grained folding energy for each sequence to input a finite-chain Ising model. The folding mechanism of the sequence is computed using a Monte Carlo simulation.*

## Simulation results

We firstly present a Monte Carlo folding simulation for *E. Coli Dihydrofolate Reductase* (EcDHFR), a model system that has been used for studying protein folding for decades [31], [32], [33]. We divided its sequence into 8 folding elements corresponding to the Minimal Common Exons previously computed for this family [25] and we assigned the internal and interaction energies for this particular protein sequence. Running simulations of the resulting finite Ising system at different temperatures, we obtained the thermal unfolding curve for each foldon. As some of the elements transition at the same temperature, only three different curves can be distinguished in Fig. 2A. As a whole, we found a complex folding mechanism with multiple steps (Fig. 2A, black dashed line). Noteworthy, it has been reported that EcDHFR folds populating several partially folded states [34], [35].

According to their folding temperature, we clustered the foldons into three separate apparent domains that are indicated in the matrix of Fig. 2D. The most stable domain, the folding nucleus, is formed by the elements 1–3 and 5–6 (yellow in Fig. 2). Counting the states with $Q$ folded elements, we defined free-energy profiles $\Delta F(Q)$ at different temperatures (Fig. 2C). The folding of the nucleus domain presents the highest free energy barrier. Once the nucleus folded, elements 7–8 fold in a subsequent step. Finally, element 4 is the less stable element, or simply the first one to unfold.

To compare our results with a traditional simulation method, we performed a structure-based (SB) model folding simulation for EcDHFR. Implementation details are included in Supplemental Methods. Keeping the foldon assignment of the simple model, we computed the folding propensity of each foldon within the defined folded ensemble of conformations (Fig. S1). The correlation with the element folding temperature of the Ising model is strong (r=0.88, p value = 0.0036). Remarkably, the relative foldability of these elements also match a reported atomistic molecular Dynamics (MD) simulation in [33], where the most flexible region matches element 4, and the other fluctuating zones belong to elements 7–8 and 1–3 according to our definition.

As a measure of the folding cooperativity, we define the cooperativity score $\rho = Q_{barrier} / (N - 1)$, the fraction of intermediary $Q$ that were not a minimum of $\Delta F(Q)$ for any *T* in a protein with *N* elements. In this case $\rho = 5/7$, lower than the cooperativity for a two-state system with a single barrier ($\rho = 1$) and higher than the cooperativity for a downhill mechanism where elements independently unfold one by one ($\rho = 0$).

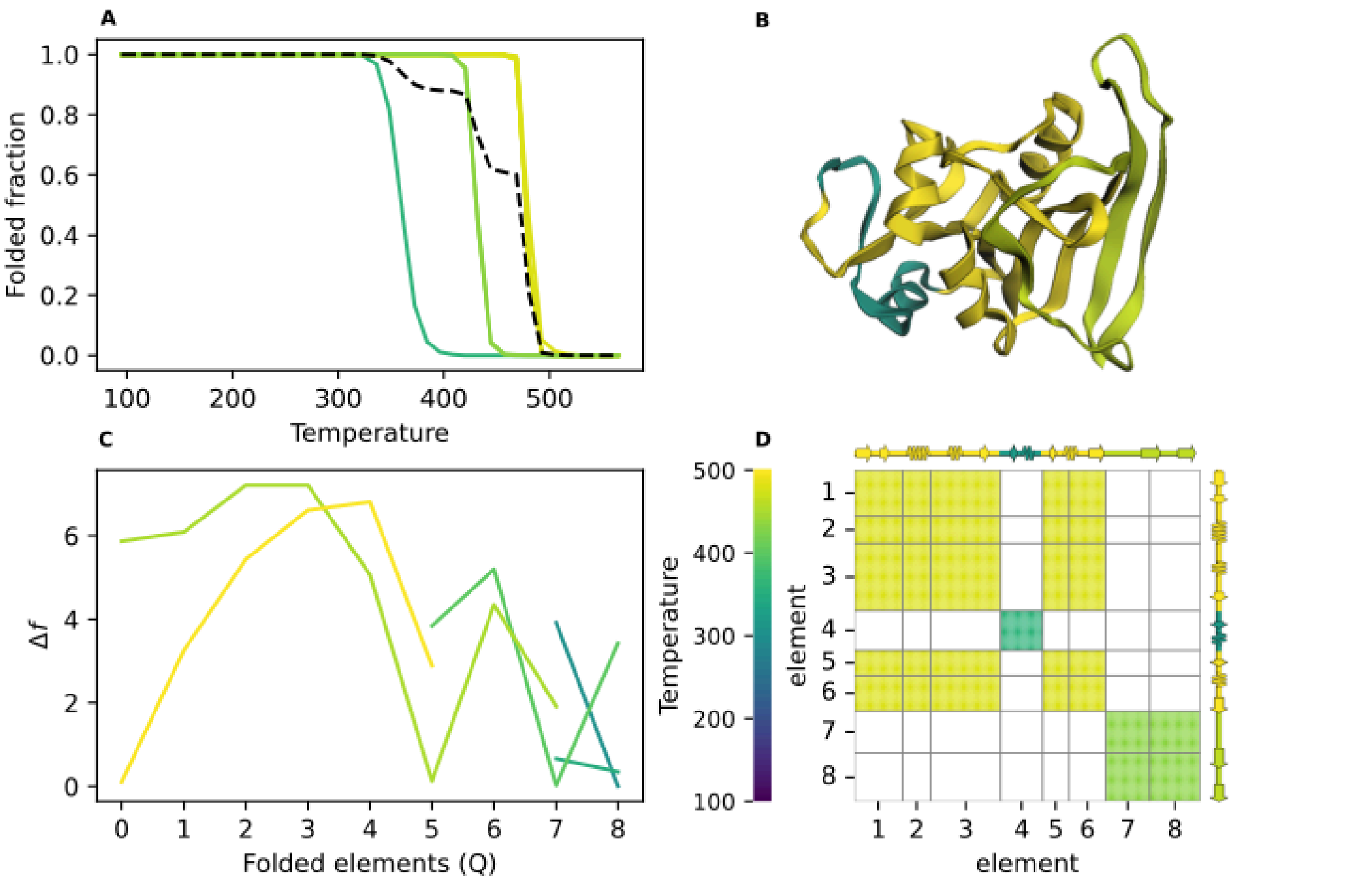

***Figure 2. Simulation results for EcDHFR .*** *(A) Simulated thermal unfolding curves, for the complete protein (black dashed line) and for each element with solid lines, colors identify the folding temperature of each one (same as B, C and D). Yellow elements are the most stable ones. (B) The structure (PDB ID 7DFR) is colored according to the folding temperature of each element. (C) Free-energy profiles, colored by temperature, with the number of folded elements Q as reaction coordinate. (D) Apparent domain matrix and secondary structure colored by element folding temperatures. The first domain to fold is formed by the elements 1, 2, 3, 5 and 6).*

We applied our simulation procedure to 15 model protein families with diverse topologies and sizes (see Table S1). The simulation results for the reference sequence of each family are included in Fig. S2. The majority of the considered proteins -9 of 15- presents a fully cooperative two-state mechanism ($\rho = 1$), while the rest, including DHFR, present at least 2 free-energy barriers and separated domains ($\rho < 1$). Interestingly, cytochromes, which require heme binding to achieve folding, exhibit the least cooperative folding behavior. The analysis of variations across different sequences within the same family is detailed in the next section.

We tested the impact of the interaction terms $\epsilon_{kj}^{s}$ in our model by repeating the simulations without including them, preserving only the internal contributions $\epsilon_{j}^{i}$. For all the studied families, in these control simulations the Cooperativity $\rho$ is significantly lower than in the Full Model (Fig. S3). Nevertheless, we identified that the families ACBP, Serpin, and Ubiquitin are still quite cooperative ($\rho > 0.6$) for $\epsilon_{kj}^{s} = 0$. Therefore, the common behaviour between their foldons is not mainly given by the strength of the interactions between them, revealing that cooperativity is only apparent, and is a consequence of the similar internal folding energies $\epsilon_{j}^{i}$.

Finally, using a control group of 10 alternative foldon partitions for each protein instead of the MCEs (Fig. S4), we found that the results are robust to alternative definitions of foldons.

*Folding mechanism variability*

For each Multiple Sequence Alignment (MSA) we obtained a single Evolutionary Potts model and common folding elements positions (the MCEs). Nevertheless, the value of the folding free-energy terms in the Ising model and therefore the predicted folding mechanism depend on each particular sequence. Can the folding mechanism be tuned by sequence modifications? How relevant are the mechanistic variations within a protein family? We analyzed 500 sequences of each family and we quantitatively analyzed the folding mechanism using two observables, the protein folding temperature $T_f$ (see Methods) and the cooperativity score $\rho$.

For each family, $T_f$ is correlated with the total evolutionary energy of the sequences (Fig. S5). This general relationship between folding stability and sequence probability is expected from Equation 2 and it is consistent with experimental results [36]. On average, the Glycolytic family is the most stable one and Trypsin the least (Fig. 3A). We note that within each family, the reference sequence (Table S1) is usually more stable than the average.

Within each family, we observed that variations in $T_f$ increase roughly linearly with the Selection Temperature, $T_{sel}$, regardless of the average $T_f$ (Fig. 3B). Consequently, families selected at lower temperatures, such as TIM or Glycolytic, only permit natural sequences with a $T_f$ close to the family average, irrespectively of the average value. This observation aligns with Miyazawa's estimation rule for $T_{sel}$ [30].

Taking advantage of the theoretical relationship $1/T_g^2 + 1/T_f^2 = 2/(T_{sel}T_f)$ [5], for each studied protein family we used the average $T_f$ and $T_{sel}$, to obtain a value for the glass transition temperature $T_g$. This allows us to calculate the ratio $T_f/T_g$, a quantitative measure of the degree of funnelness of a folding landscape, with higher values corresponding to more ideal funnels, and also $T_{sel}/T_g$ which quantify the degree of evolutionary optimization, with lower values corresponding to more highly optimized sequences [8], [37]. All the studied families present Temperature ratios within the energy landscape theoretical limits, $T_f/T_g > 1$ guaranteeing that they can fold in relevant timescales and $T_{sel}/T_g < 1$, ensuring that they can evolve in a relevant timescale (Fig. 3C). Over both theoretical limits, Trypsin family has the least optimized sequences and the less funneled landscape, allowing large variations in $T_f$ given by sequence modifications. Glycolytic family, which we identified as exhibiting the lower $T_f$ variations from the average, has the most foldable and evolutionarily constrained sequences.

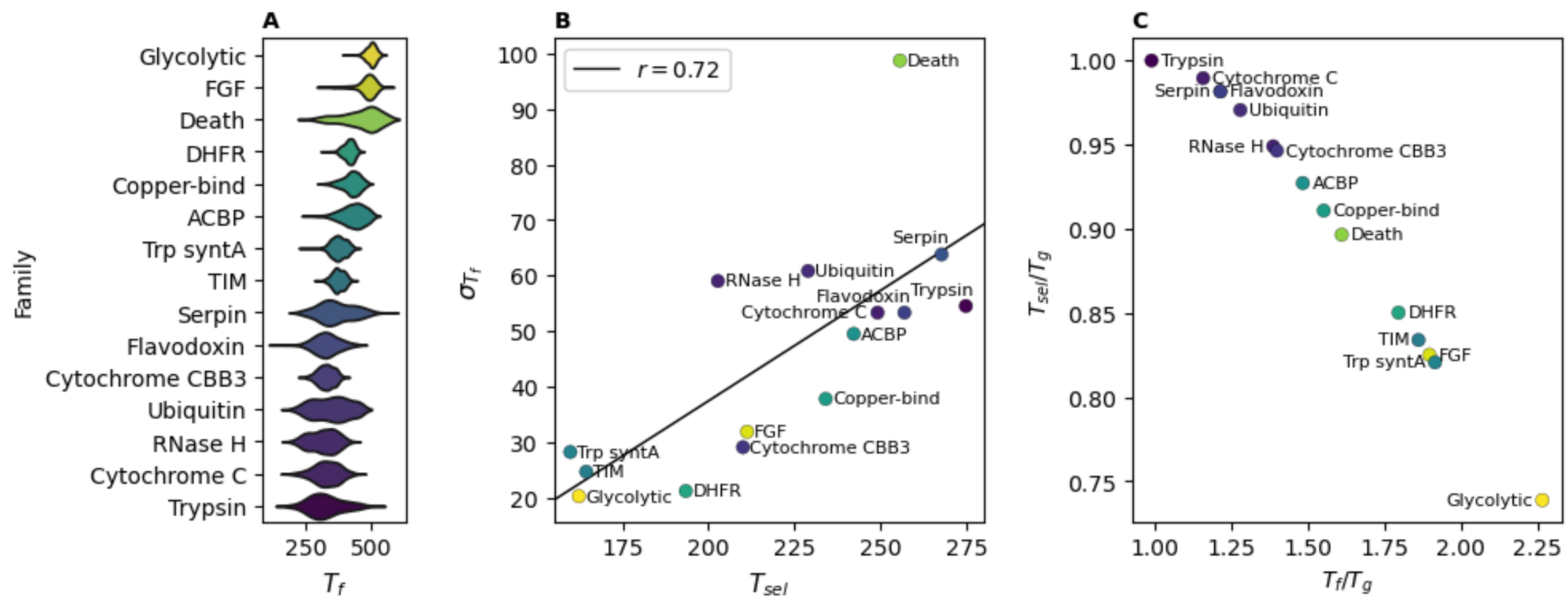

***Figure 3**. **Variation of the Folding temperature.** (A) Distribution of the protein Folding Temperature ($T_f$) within each family with a color scale according to the family-average. (B) Correlation between the standard deviation of the protein folding temperature and the Selection Temperature ($T_{sel}$), The color scale is the same as that in panel A. (C) Selection Temperature to Glass transition Temperature ratio ($T_{sel}/T_g$) versus the family-average Folding Temperature to Glass transition Temperature ratio ($T_f/T_g$) for all the studied protein families. The color scale is the same as that in panel A.*

Analyzing the average folding domains of multiple sequences per family reveals that the two-state behaviour registered for the majority of the reference sequences is not widespread within all those families (Fig. S6). This is for instance the case of the Ubiquitin family, where the folding cooperativity varies notably among family members.

We found that the cooperativity is also strongly related to the sequence evolutionary energy. The folding mechanism becomes more cooperative for sequences with folding elements that have comparable energy and a stronger coupling between them. Moreover, the cooperativity score smoothly changes in a parameter space given by the energetic heterogeneity of foldons and their average interaction strength (see Methods). This trend is consistently observed across all studied protein families (Fig. 4A, B and Fig. S7). Interestingly, for some families such as DHFR, cooperativity within a narrow range and their natural sequences concentrate in a specific region of the heterogeneity-interaction space (Fig. 4A). For other families such as ACBP, the folding mechanism varies from all-or-none transitions ($\rho = 1$) to completely downhill mechanisms ($\rho = 0$). For this latter family, sequences are distributed across the heterogeneity-interaction space (Fig. 4B).

The difference between these two families is not merely anecdotal, but follows a general trend evidenced in the plot of Fig. 4C. The variance of the Cooperativity Score is correlated to the ratio between the number of short-range and long-range native contacts $N_{Short} / N_{Long}$ (see Methods). This is in line with the findings of *Cho et al.* about the topological sensitivity of the transition state fluctuations in two-state proteins [38]. The ACBP family and the other $\alpha$-proteins with an elongated architecture (see Table S1) allow natural sequences with a variety of folding mechanisms. For some $\beta$ and $\alpha/\beta$ structures, such as DHFR, more compact and with more short-range native contacts, the topology appears to restrict the possible mechanisms. Only one $\beta$ family, Copper-bind, seems to escape the trend (Fig. 4C). We highlight that there are families such as Serpin, Flavodoxin and Trypsin that present a high $T_{sel}$, and consequently a high total energetic and sequence variability, but a very conserved cooperativity score (Fig. S8) as expected by their low $N_{Short}/N_{Long}$.

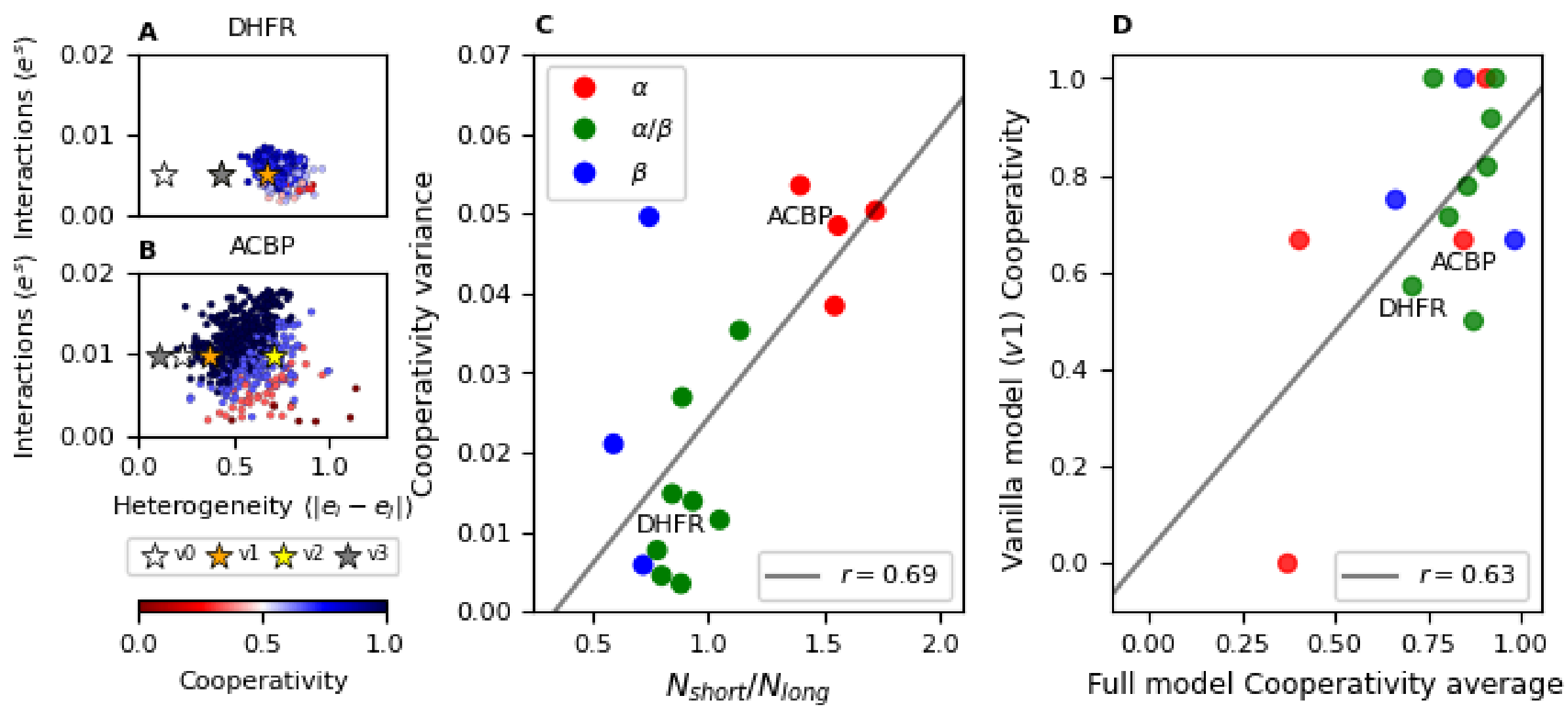

***Figure 4**. **Variation of the Cooperativity.** (A) Cooperativity score for 500 representative sequences of DHFR family is shown in a color scale on a plane defined by the energetic heterogeneity of foldons and their average interaction strength (see Methods). Vanilla models are marked as stars. The v1 model (orange star) has the closest to natural sequence heterogeneity and interactions. (B) Sequences of the ACBP family, in the same space and cooperativity color scale as that in panel A. (C) Correlation between the cooperativity score variance for each family and the rate between the number short-range and long-range contacts in the reference PDB structure (see Methods). The folding mechanisms are more diverse within Alpha protein families (such as ACBP, red dots) than within alpha/beta or beta proteins (green and blue dots). (D) Correlation between the cooperativity obtained with the vanilla model v1 and, on average, with the full model for each family. The color code is the same as that in panel C.*

Given that for some protein families the folding mechanism that the Ising model predicts is conserved regardless of the sequence, we compare our results with alternative perfect funnel 'vanilla' models that are not sensitive to the specific amino acid identities. We test the same Ising coarse-grained model but replacing the evolutionary Potts fields with constant energy fields for all the sequences of the same family. Therefore, to calculate folding energy terms, we replaced in Equation 2 the 'flavoured' evolutionary couplings $J_{ab}(\sigma_a, \sigma_b)$ with the 'vanilla' contact matrix $C_{ab}$. Also, we replaced the 'flavoured' local field $h_a(\sigma_a)$ with a binarized secondary structure element assignment $h_a^1 = \delta_{\alpha/\beta,\, coil}$ (where $\delta = 1$ if a position is $\alpha$ or $\beta$ and $\delta = 0$ if it is not). We call this model *v1.* As the choice for a vanilla local field $h_a$ is not obvious, we implemented several alternatives (see models *v0, v2*, and *v3* in the Supplemental Information). The Potts-like v1 model is purely topological and it only uses a common single reference structure for every protein of each family. However, according to Equation 2, the resulting Ising Hamiltonian is still heterogeneous, with energy terms varying according to the $h_a$ average per foldon and the contact counting within and between foldons. We normalize the vanilla fields such that $\sum C_{ab}$ = $<\sum J_{ab}(\sigma_a, \sigma_b)>$ and $\sum h_a^1 = <\sum h_a(\sigma_a)>$, where <*> is the average over natural sequences. Therefore, by construction the protein folding temperature with the vanilla model would match the average of the corresponding family. There is a correlation (r = 0.63) between the cooperativity score values obtained with the vanilla model and the one obtained with the full, flavored Ising model (Fig. 4D and Fig. S9). Consistently, in the energetic heterogeneity-interactions space (Fig. 4A, B and Fig. S7), the vanilla model is close to natural sequences. For families such as Glycolytic or Serpin, the variations in cooperativity are so limited that a purely topological vanilla model like *v1* can be a reasonable approximation. However, for the $\alpha$-proteins, where the $N_{Short}$ / $N_{Long}$ ratio is higher, a flavoured model that takes into account each particular amino acid sequence is needed to determine the folding mechanism.

*Prediction of mutational effects on the folding mechanisms*

The stability and cooperativity of the folding of natural proteins can be perturbed both by environmental factors and by sequence modifications. In general it is experimentally observed that single point mutations destabilize the native states by a few kcal/mol, and many computational approximations are being constructed to capture these effects. Notably, the simple folding energetic model that we propose here captures the experimentally observed change in temperature denaturation for many protein families (Supplemental Text, Fig. S13 and Table S2). Moreover, we found that the Cooperativity Score ρ for different proteins are well correlated with the experimental m-values determined by chemical denaturation (Fig. S12, r = 0.74, p value = 3.62e-80). Both the folding temperature $T_f$ and the cooperativity score ρ are directly related to the sequence evolutionary energy. We quantified this relationship with a linear fit of ρ for the natural sequences of all the studied protein families (Fig. 5A) and family-specific linear fits for $T_f$ (Fig. 5B-C). Given a sequence and without running any additional Monte Carlo simulation, stability and cooperativity changes ($\Delta T_f$ and Δρ) upon possible single-site mutations can be estimated. We present the $\Delta T_f$ and Δρ predictions for all possible single-site mutations of DHFR (Fig. 5D-E) and ACBP (Fig. 5F-G). For each family, we chose the closest natural sequence to the ρ and $T_f$ family-average as the wild type.

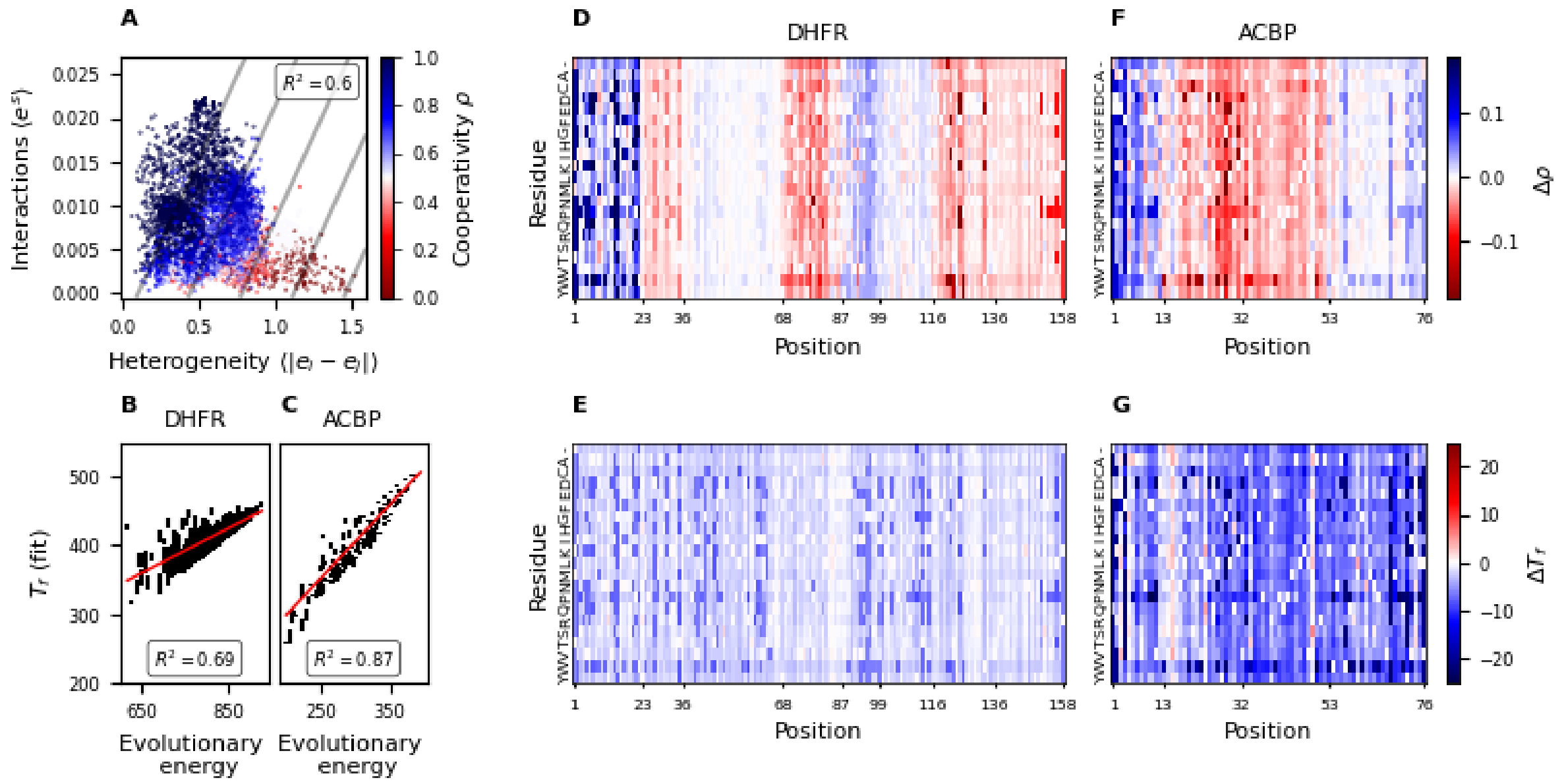

***Figure 5*. *Single site mutant predictions*.** *(A) Cooperativity scores of all simulated sequences, for all families in a color scale on the space defined by the energetic heterogeneity of foldons and their average interaction strength (see Methods). The grey lines are level curves, given by a linear fit. (B) Protein Folding Temperature fitted for each protein vs Evolutionary Energy for the natural sequences of DHFR family. A linear fit is shown in red. (C) Same relationship as that in panel B, for ACBP family. (D) Changes in the Folding Temperature upon all single-point mutations, for a natural DHFR sequence. (E) Changes in the Folding Temperature upon all single point mutations, for a natural ACBP sequence. (F) Changes in the Cooperativity score upon single point mutations, for a natural DHFR sequence. (G) Changes in the Cooperativity score upon single point mutations, for a natural ACBP sequence.*

For both the studied proteins (Fig. 5F and G), the $\Delta T_f$ results show that the majority of point mutations destabilize the structure. Nevertheless, these natural proteins are not occupying exactly a local minima in the sequence space, because there are some possible stabilizing mutations. For both examples, there are sites where, depending on the specific amino acid choice, the $T_f$ of the protein can significantly increase or decrease. Folding temperature changes upon single site mutations are on average larger for the ACBP sequence (Fig. 5G) than for the DHFR sequence (Fig. 5E), as happened for the variations across different natural sequences within each of these two families (Fig. 3). However, we did not identify the same behavior across all families (Fig. S10), evidencing epistatic effects.

Although the global stability predictions could be made directly applying the Evolutionary Model and considering the family $T_{sel}$, the power of the proposed framework is to quantify how do point mutations affect the Cooperativity Score ρ (Fig. 5D and F). The folding elements assignment modulates the effect of local stability changes. For instance, the destabilizing mutations (blue regions in Fig. 5E and G) in highly stable elements reduce the energetic heterogeneity, producing a more cooperative folding (red vertical stripes in Fig. 5D and F). On top of that, the mutation impact on inter-element interactions also affects Δρ predictions. Finally, the cooperativity variance of natural sequences presents a positive correlation (*r = 0.63*) with the predicted cooperativity variance for point mutations (Fig. S11).

Therefore, it is expected that families like ACBP and other $\alpha$-proteins to be the most sensitive to engineer single variants with different folding cooperativity.

## Concluding remarks

Protein folding and evolution are two deeply intertwined phenomena. On the one hand, polypeptides must be able to fold in physiological time-scales, a fact fundamentally constrained by the physico-chemical nature of their folding landscape. On the other hand, the sequences must be able to change in evolutionary time-scales, allowing for structural variations to be explored in their evolutionary landscapes. We presented here a quantification of the relation between both landscapes by exploring the folding mechanisms variations given by sequence modifications.

The proposed scheme to map the evolutionary energy to a coarse-grained folding model readily allows for the calculation of the equilibrium temperature-induced denaturation, the free energy profile and the emergence of subdomains for any sequence of a given protein family (Fig. 2). For the majority of the reference sequences for each family, the folding appears as two-state, populating the fully folded and fully denatured states (Fig. S2). This is in line with the experimental findings of these well-behaved folding protein models [39]. However, we note that in the families that present higher folding variability (Fig. S7), the reference sequence may not be the archetypical case. The effective cooperativity is brought about by the interaction energy terms between foldons (Fig. S3), and is not a consequence of the similarity of the internal energetic terms of each foldon that could give rise to an apparent overall cooperativity.

The mean folding temperature $T_f$ for each family is related to the funneling of the folding energy landscape given by the $T_f/T_g$ ratio (Fig. 3). This constrains the $T_{sel}/T_g$ ratio for each family, and thus the impact of natural sequence variants to the change of $T_f$ within each family can be approximated by their respective $T_{sel}$ (Fig. 3). In other words, the lower the $T_{sel}$, the less sequences can be evolutionarily selected that can match the needed $T_f$. It is worth noting that this allows for an immediate annotation of expected folding temperature for any natural sequence of the family, which can be readily experimentally tested.

The mean folding cooperativity for each family can be explained by purely topological models of the native state, as expected. We found that a simple assignment of presence/absence of regular secondary structural elements to the local fields together with a native contact based potential is enough to explain the mean cooperative behavior for all families (Fig. 4). However, to capture the variations in the cooperativity given by sequence modifications a full evolutionary informed model has to be applied. We found that the variation in the cooperativity can be reasonably approximated by the ratio between the number of short-range and long-range native contacts, distinguishing mainly $\alpha$ , $\beta$ or $\alpha/\beta$ overall topologies. Interestingly, there are families where a high energetic and sequence diversity is allowed by an elevated $T_{sel}$ (Fig. S8), but this does not impact on any mechanism variability because of the topological restrictions (Fig. 4).

Taking the folding simulations of 7500 proteins of all families together, we can see that the folding cooperativity can be very well predicted by the energetic variations within and between foldons (Fig. 5). High cooperativity results from sequences that have low between-foldon energetic variation and high inter-foldon interaction energy, whereas low cooperativity is explained in the opposite scenario. Notably, no natural sequence was found to occupy regions of this space where both parameters are high, coinciding with the results found for repeat-proteins [26]. This allows for the prediction of the effect of mutations on both the folding temperature and the folding cooperativity without doing the folding simulation (Fig. 5). This is equivalent to the prediction of the effect of every possible single-site mutant for any protein sequence. We have found that this simple calculation captures the effect of mutations in various protein families for which experimental data is available (Supplemental Text and Fig. S13). At present, the proposed model is applicable for protein systems for which a meaningful multiple sequence alignment can be built such that local energetics can be estimated. This leaves out intrinsically disordered proteins and low complexity regions of natural proteins. For natural large multi-domains proteins defining the interaction energetics between domains is a difficult issue that at this point must be treated in a case by case manner.

Unfortunately, there is currently no high throughput technology to measure the folding mechanisms with precision in the laboratory. Considering this limitation, we developed a local and coarse-grained mapping of the sequence probability distribution to folding stability, allowing a computational exploration of the folding mechanism for any protein family for which sufficient sequence data is known. We highlight that this framework assumes that folding stability is locally the main evolutionary constraint, an approximation in line with the minimal frustration principle. Therefore, sequence positions strongly conserved and conditioned by other selection forces besides folding may affect local stability and some cooperativity predictions, locally frustrating the folding landscape [40]. On the contrary, the folding variability among protein family members will stand irrespective of the local frustration brought about by other selective pressures. We want to emphasize that the proposed model provides a reproducible way to rank natural sequence variants in terms of their folding stability, cooperativity, or both, which can be harnessed for finding useful homologs for biotechnological applications.

## Methods

*Data curation*

We used a total of 15 well behaved protein families with distinct topologies (see Table S1). We obtained the Multiple Sequence Alignments (MSA) from Pfam [41], now hosted by INTERPRO database [42] (consulted in December 2022). Additionally, we considered an MSA for PDZ family for computing the reference Selection Temperature (see below). We used a reference PDB for each family (Table S1) and we aligned the MSA to its corresponding sequence, keeping only the positions of the MSA that are present in the target sequence. To summarize, MSA positions are Pfam domain positions in the target PDB structure. For minimizing the phylogenetic bias within each MSA, we clustered by full sequence similarity using CD-hit [43] at 90% cutoff and we assigned a weight to each sequence defined as $1/n_i$ being $n_i$ the number of sequences in the $i$th cluster. All the statistics were made taking into account these sequence weights.

*Restricted Boltzmann Machine*

We applied the Restricted Boltzmann Machine (RBM) method developed by Tubiana et al. [28]. The model has two layers, a visible layer is given by the MSA positions and a hidden layer. Interactions are not allowed between the layers, but only within them. We used a quadratic hidden-unit potential (Gaussian RBM) for which the learned weights were exactly mapped to the effective pairwise couplings between the visible units, the Potts model parameters [28]. We tested the performance of the RBM learning for a grid of parameters for the DHFR family and for all cases we decided to use 500 iterations and 500 hidden units, with a regularization strength of $\lambda_1^2 = 0.25$. For the learning we imposed a gap threshold on the MSA, removing sequences with more than 10% of gaps, except for the families Cytochrome CBB3, Flavodoxin. ACBP and Ubiquitin, where relaxing the threshold to 70% of gaps increased the model likelihood.

*Minimal Common Exons*

We used the Minimal Common Exon (MCE) for each protein family, a sequence partition given by the exon boundary hotspots of the Family, as defined in ref [25]. For controls, we used alternative partitions, for which the element size was sampled for a neutral distribution given by the natural MCEs, as defined in ref [25].

*Selection Temperature*

We calculated the Selection Temperature $T_{sel}^{PDZ}$ for a reference family, PDZ comparing the experimental $\Delta\Delta G$ data [44] with the corresponding Evolutionary Energy differences $\Delta E$ obtained with a RBM method (Fig. S14). Using $T_{sel}^{PDZ}$, we estimate the Selection Temperature $T_{sel}$ for the studied families as

$$T_{sel}\, \sigma\, (\Delta E) = T_{sel}^{PDZ}\, \sigma\, (\Delta E^{PDZ})\ , \tag{3}$$

where $\sigma\,(\Delta E)$ denotes the standard deviation of Evolutionary Energy differences upon point mutations averaged over homologous sequences [30]. The results are provided in Table S1.

*Folding Ising Model Monte-Carlo Simulations*

We performed Monte Carlo Metropolis algorithm simulations of the finite Ising model with a python routine. The code is available at GitHub (https://github.com/eagalpern/folding-ising-globular). Simulation total time, transient time and equilibration time parameters were obtained with an autocorrelation analysis.

*Free energy profiles*

We obtained free energy profiles approximating the probability of states $s$ with $Q$ folded elements with the Metropolis Monte-Carlo sampling. We considered together sampled states for simulations performed in a window of the 10 closest temperatures. The profiles we used are computed as

$$\Delta f\,(Q) = -k_B T\, log\left(\frac{\sum_{s|Q} N(s)}{\sum_{s} N(s)}\right), \tag{4}$$

where $T$ is the average temperature, $N(s)$ are the counts of state $s$ and $s|Q$ are the states with $Q$ folded elements.

*Cooperativity Score*

*The Cooperativity Score* ρ *is defined from the free energy profiles* $\Delta f\,(Q)$ *obtained for the protein, where the reaction coordinate Q is the number of folded elements. At each temperature, there is a Q were* $\Delta f\,(Q)$ reach its minimum, therefore the stable state has Q folded elements. Taking together all the free energy profiles, there could be some values of Q where there is never a minimum of $\Delta f\,(Q)$, but a free energy barrier. Excluding the completely folded and unfolded system, the Cooperativity Score is defined by counting how many of the $N - 1$ possible intermediary values of Q that are never a free energy minimum,

$$\rho = Q_{barrier} / (N - 1). \quad (5)$$

*Apparent domains*

Elements *j, k* were assigned to the same domain if $|T_f^j - T_f^k| < 5$, where the folding temperature $T_f^j$ was obtained by a sigmoid fit of the folding probability of element *j*. Overlapping domains were separated into the minimum number of non-overlapping ones. If more than one separation is possible, temperature differences between domains are maximized.

*Folding Temperature*

To fit folding temperatures $T_f^j$ we approximated the fraction folded $m(T)$ as

$$m(T) = \frac{m_{max}}{1 + e^{a(T - T_f)}}, \quad (6)$$

where $m_{max} \in [0, 1]$. We used the *scipy* library *curve_fit* to fit $T_f$ and to get the standard deviation that we used as $T_f$ errors.

*Foldon Energetic Heterogeneity and Interactions*

We defined the foldon energetic heterogeneity as the average internal energy difference between foldons $< |e_k^i - e_j^i| >$, where $<*>$ indicates an average over all *j,k* with *j>k* protein folding elements. We used a length-normalized energy to calculate differences, $e_k^i = \epsilon_k^i / L_k$, where $\epsilon_k^i$ is defined in Equation 2 and $L_k$ is the sequence length of the folding unit. For the energy interaction strength we used the average normalized surface energies $< e_{jk}^s >$, where $e_{jk}^s = \epsilon_{jk}^s / (L_j L_k)$.

*Short and long-range contacts*

For each family, we calculated a contact map from the reference PDB (Table S1) using a distance threshold of 9.5Å between C-betas (and C-alpha in the case of Glycine). We defined as long-range the contacts *i, j* such that | *i* - *j* | ≥ 6, and short-range the contacts *i, j* such that 1 < | *i* - *j* | < 6.

## Data Availability

Data has been deposited in GitHub (https://github.com/eagalpern/folding-ising-globular).

## Code Availability

Code has been deposited in GitHub (https://github.com/eagalpern/folding-ising-globular).

**Author Contributions.** EAG and EAR performed research, EAG, EAR and DUF designed research, analyzed data, and wrote the manuscript**.**

## Acknowledgements

We thank Ignacio E. Sánchez for his valuable insights and constructive dialogue throughout the course of this research. This work was supported by the Consejo de Investigaciones Científicas y Técnicas (CONICET) (EAR and DUF are CONICET researchers and EAG is a postdoctoral fellow); CONICET Grant PIP2022-2024—11220210100704CO and Universidad de Buenos Aires grant UBACyT 20020220200106BA. We call the attention of the international scientific community about the catastrophic erosion of Argentina's strong scientific tradition due to current funding constraints and the sudden termination of long term policies.

**Supplemental Information for**

*Inferring protein folding mechanisms from natural sequence diversity*

Ezequiel A. Galpern, Ernesto A. Roman and Diego U. Ferreiro*

* To whom correspondence may be addressed: ferreiro@qb.fcen.uba.ar

# Supplemental Text

## Supplemental Methods

**Structure-based model simulations.** We performed single-basin Cα-Structure-based model simulations (Cα-SBM) using native bonds, angles and dihedrals corresponding to PDB ID 7DFR of Dihydrofolate reductase enzyme. Each atom is represented as a single bead of unit mass where bond lengths, angles, and dihedrals are maintained by harmonic potentials. Non-bonded atom pairs that are in contact in the native state between residues *i* and *j*, where *i* > *j+3*, are given a Lennard-Jones potential, while all other non-native interactions are repulsive. The contacts are defined using the Shadow Map algorithm using default parameters [1].We performed simulations around the folding temperature with specific native contacts biases using umbrella sampling technique. We combined them using the Weighted Histogram Analysis Method (WHAM) [2] to calculate the free energy profiles. For each residue in the protein structure one can calculate the folded state (Qi) which is the proportion of native contacts formed in state i with respect to the total number of native contacts. In this way, Qi goes from 0 to 1 and indicates the nativeness of that residue. In our work, we grouped sequential segments of residues defined previously as protein foldons and, instead of calculating the nativeness of each residue, we calculated the nativeness for a given foldon (*Foldon Qi*), which is the average of Qi values for the residues in the foldon. To calculate folding propensity, we take the mean value of *Foldon Qi* within the folded ensemble, which extends within a boundary of 2 kcal/mol from the corresponding Free energy minimum shown in Figure Fig. S1A.

**Vanilla Ising Models.** We defined perfect funnel 'vanilla' models that are not sensitive to the specific amino acid identities, replacing the evolutionary Potts fields with constant energy fields for all the sequences of the same family. The couplings $J_{ab}(\sigma_a, \sigma_b)$ were replaced with the 'vanilla' contact matrix $C_{ab}$corresponding to the reference PDB structure. The choice for a vanilla local field $h_a(\sigma_a)$ is not obvious, hence we propose several vanilla candidates, starting with a trivial uniform value $h^0$. We call this model *v0*. Alternatively, we define model *v1* using a binarized secondary structure element assignment $h_a^1 = \delta_{\alpha/\beta, coil}$ (where $\delta = 1$ if α or β and $\delta = 0$ if it is not), a model *v2* using the Relative Accessible Surface Area (RASA) as $h_a^2 =- log(RASA)$ and v3 using the information content per position $h_a^3 = log_2(20) - H_a$, where $H$ is the Shannon entropy.

Except for *v3*, where the conservation is obtained from the family MSA, these proposed vanilla Potts-like models are purely topological. According to Eq. 2, the resulting Ising Hamiltonian is still heterogeneous, with energy terms varying according to the $h_a$ average per foldon and the contact counting within and between foldons. We normalize the vanilla fields such that $\sum C_{ab} = <\sum J_{ab}(\sigma_a, \sigma_b)>$ and $\sum h_a^x = <\sum h_a(\sigma_a)>$, where $<*>$ is the average over natural sequences. Therefore, by construction the protein folding temperature with any vanilla model would match the average of the corresponding family.

**Comparison with experimental data.** We validate the results of our model with extensive literature data. We compare our predictions with the experimental data available in the ProTherm database [3]. We only consider reversible folding transitions and we keep a single wild type sequence per family. Experimental folding cooperativity is defined by the ΔG dependence on a denaturing agent, such as a chaotrope, and is reported as the m-value. Given that Protherm contains this information we studied the correlation between this experimental observation with the cooperativity obtained with the simple foldon model we propose. We found that the Cooperativity Score values that we predicted for different proteins, including single site mutants, are correlated with experimental m-values (Fig. S12, r = 0.74, p value = 3.62e-80). We can conclude that, with the exception of the CytochromeC family, the Cooperativity score that we introduce and compute with a simple foldon model is well correlated with the experimentally reported m-values. We note that CytochromeC is an heme-containing protein, and our model does not contain any energetic term to account for the cofactor interaction, so it may be not surprising that the model fails to capture the experimentally observed cooperativity. For the stability change predictions, we compare our simple model for the changes in the folding temperature upon single site mutants $T_f$ versus experimental values from the ProTherm database. As these may depend on other environmental factors, we separate also between the different experiments and pH conditions. To make a fair comparison, we eliminate experimental non-reversible transitions, experiment with protein fragments or in the presence of external binders. We keep experiments for which at least two different single sites were mutated and with a total of at least 10 mutants considering all sites. The results are summarized in Fig. S13 and Table S2. Overall it can be seen that the experimentally measured change in folding stability of the mutants is reasonably well predicted by our simple model. Interestingly, in the cases where active sites were mutated to stabilize the wild type protein, the correlation does not hold, such as the Rnase H family. It is important to highlight here that our model is built on the assumption that protein folding is the sole evolutionary driving force, which of course is not true in the case of enzymes, where sequence conservation of the active sites is needed for chemical activity and may even conflict with robust folding of the domain [4].

## Supplemental Tables

| Family | PDB | $\alpha$ or $\beta$ | Reference sequence | Seq Length | Pfam | Effective number of sequences | $T_{sel}$ | Foldons |
|---|---|---|---|---|---|---|---|---|
| ACBP | 2abd | $\alpha$ | P07107.2/3-78 | 76 | PF00887 | 7238 | 242 | 4 |
| Copper-bind | 5pcy | $\beta$ | P00299.2/71-168 | 98 | PF00127 | 4427 | 234 | 5 |
| Cytochrome C | 1cyc | $\alpha$ | P00025.2/4-102 | 99 | PF00034 | 14707 | 249 | 3 |
| Cytochrome CBB3 | 451c | $\alpha$ | P00099.2/25-100 | 76 | PF13442 | 75178 | 210 | 4 |
| Death | 1e3y | $\alpha$ | Q13158.1/101-179 | 79 | PF00531 | 5777 | 256 | 4 |
| DHFR | 7dfr | $\alpha/\beta$ | P0ABQ4.1/1-158 | 159 | PF00186 | 12705 | 193 | 8 |
| FGF | 1rg8 | $\beta$ | P05230.1/28-148 | 121 | PF00167 | 1693 | 211 | 4 |
| Flavodoxin | 5nll | $\alpha/\beta$ | P00322.1/3-128 | 126 | PF00258 | 23868 | 257 | 7 |
| Glycolytic | 1ado | $\alpha/\beta$ | P00883.2/15-364 | 350 | PF00274 | 2851 | 162 | 12 |
| RNaseH | 1hrh | $\alpha/\beta$ | P03366.3/1037-1155 | 119 | PF00075 | 20257 | 203 | 4 |
| Serpin | 1qlp | $\alpha/\beta$ | P01009.3/54-415 | 362 | PF00079 | 10481 | 268 | 13 |
| TIM | 1tim | $\alpha/\beta$ | P00940.2/7-246 | 240 | PF00121 | 12402 | 164 | 8 |
| Trp syntA | 1bks | $\alpha/\beta$ | P00929.1/8-266 | 259 | PF00290 | 11709 | 160 | 10 |
| Trypsin | 3ptn | $\beta$ | P00760.3/24-239 | 216 | PF00089 | 43192 | 275 | 7 |
| Ubiquitin | 1ubq | $\alpha/\beta$ | P0CG48.3/611-682 | 72 | PF00240 | 28983 | 229 | 5 |

***Table S1. Summary of the studied protein families.*** *Information on 15 families, including the reference PDB ID, the topology classification, the Uniprot ID of the reference sequence, the corresponding sequence length, the family Pfam ID, the effective size of the Multiple Sequence Alignment (MSA), the Selection Temperature and the number of folding elements. Details about data curation are available in the Methods section.*

| Family | PMID | Author | Year | pH | points | slope | intercept | r | p value |
|---|---|---|---|---|---|---|---|---|---|
| DHFR | No ID | Thompson | 2020 | 8 | 24 | 0.08 | -1.25 | 0.42 | 0.043 |
| Trp syntA | 8795042 | Hiraga K et al. | 1996 | 9 | 11 | 0.17 | -0.89 | 0.65 | 0.032 |
| DHFR | 25905910 | J. Tian et al. | 2015 | 7.8 | 16 | 0.16 | -0.51 | 0.65 | 0.0063 |
| Flavodoxin | 15504414 | Campos LA et al. | 2004 | 7 | 25 | 1.13 | -0.52 | 0.68 | 0.0002 |
| RNase H | 9228039 | Akasako A et al. | 1997 | 3.2 | 18 | 0.22 | -4.56 | 0.69 | 0.0015 |
| RNase H | 8955106 | Kanaya S et al. | 1996 | 9 | 14 | 0 | -5.7 | 0 | 0.99 |
| Cytochrome C | 7880822 | Pielak GJ et al. | 1995 | 4.6 | 11 | 0.24 | -3.28 | 0.67 | 0.023 |
| Trp syntA | 1727648 | Lim W.K. et al. | 1992 | 7.2 | 28 | 0.08 | -2.58 | 0.14 | 0.48 |

***Table S2.*** *Summary of the comparison to experimental data of folding temperature change predictions upon single site mutations.*

## Supplemental Figures

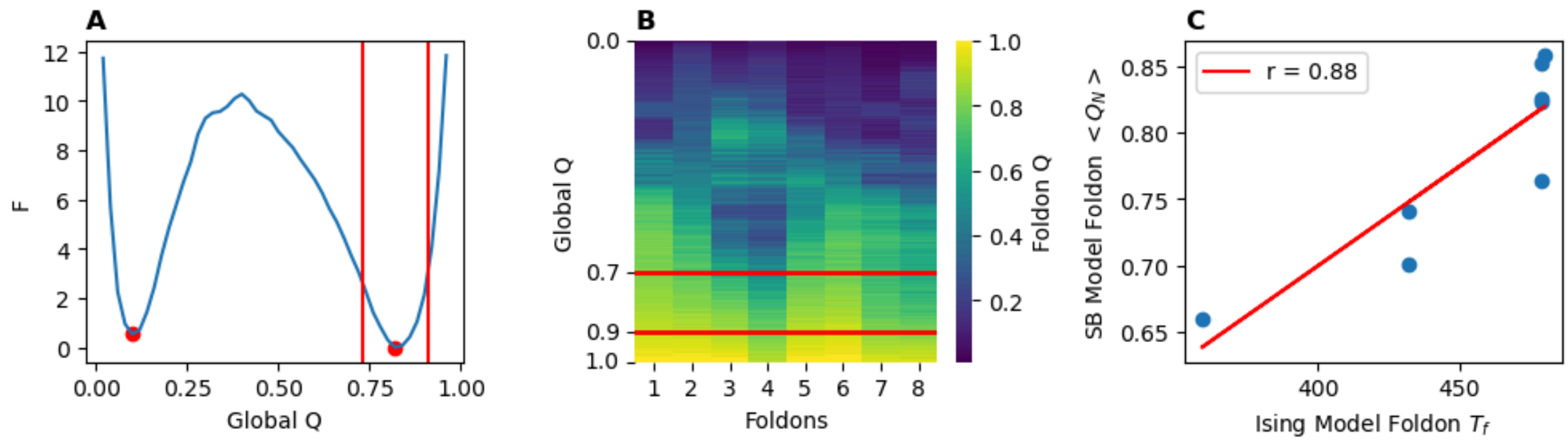

***Figure S1. Structure-based model simulation for EcDHFR.*** *(A) Free energy profile as a function of native contacts (blue). Red dots and lines are the bottom and boundaries of the unfolded and folded conformational ensemble wells, respectively. (B) Evaluation of the Element folded state (Foldon Q) at each position in the native contacts reaction coordinate (Global Q). (C) Correlation between folding propensity in the structure-based model simulation for the folded ensemble of conformations, measured as Foldon Q value, with the element folding temperature in the Ising model.*

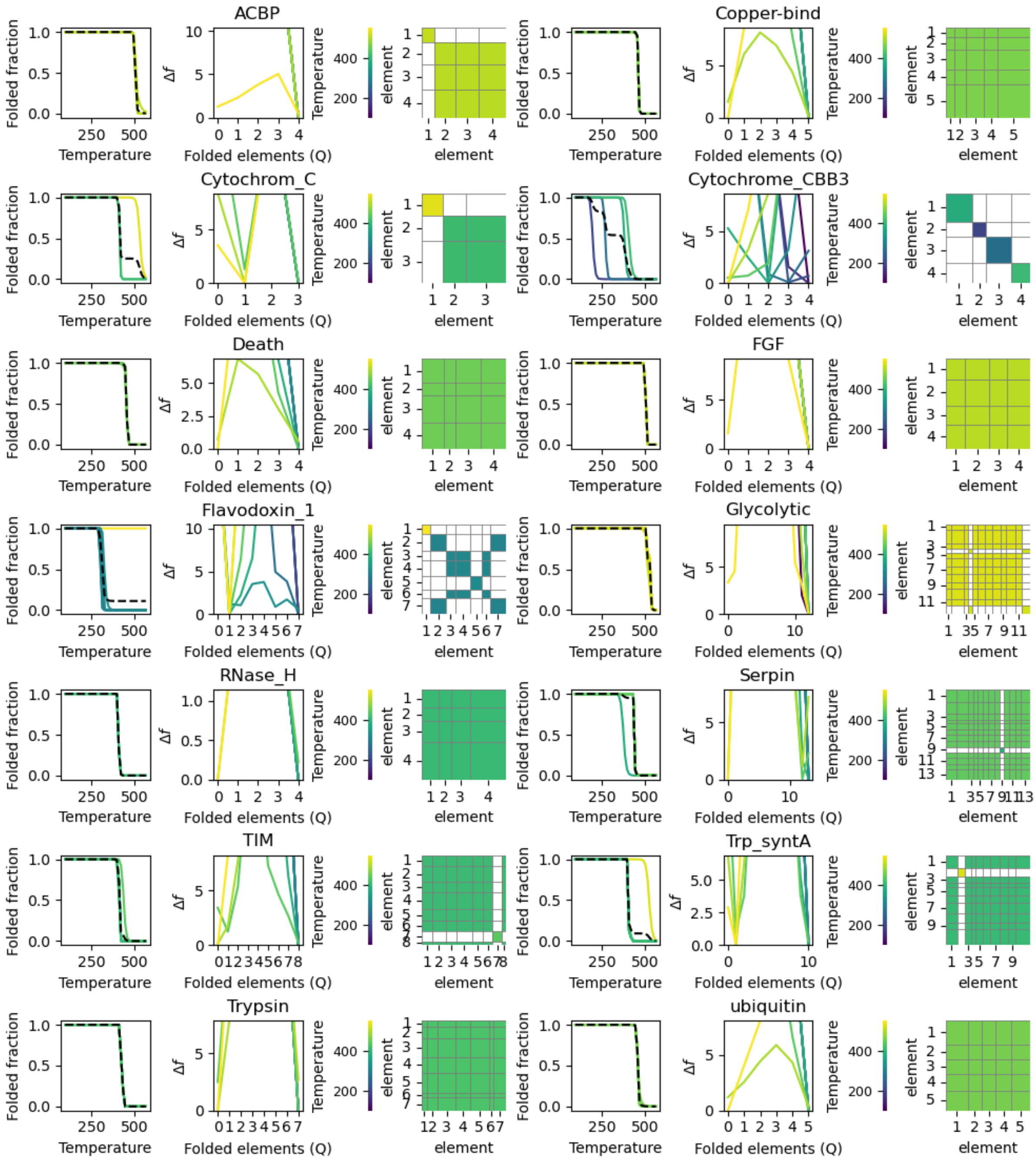

***Figure S2. Simulation results.*** *For a reference sequence of each protein family (see Table S1), from left to right panels, the simulated thermal unfolding curves, for the complete protein (black dashed line) and for each element with solid lines, colors identify the folding temperature of each one. Yellow elements are the most stable ones. The PDB structure, colored according to the folding temperature of each element. Free-energy profiles, colored by temperature, with the number of folded elements Q as reaction coordinate. Apparent domain matrix, colored by domain folding temperatures.*

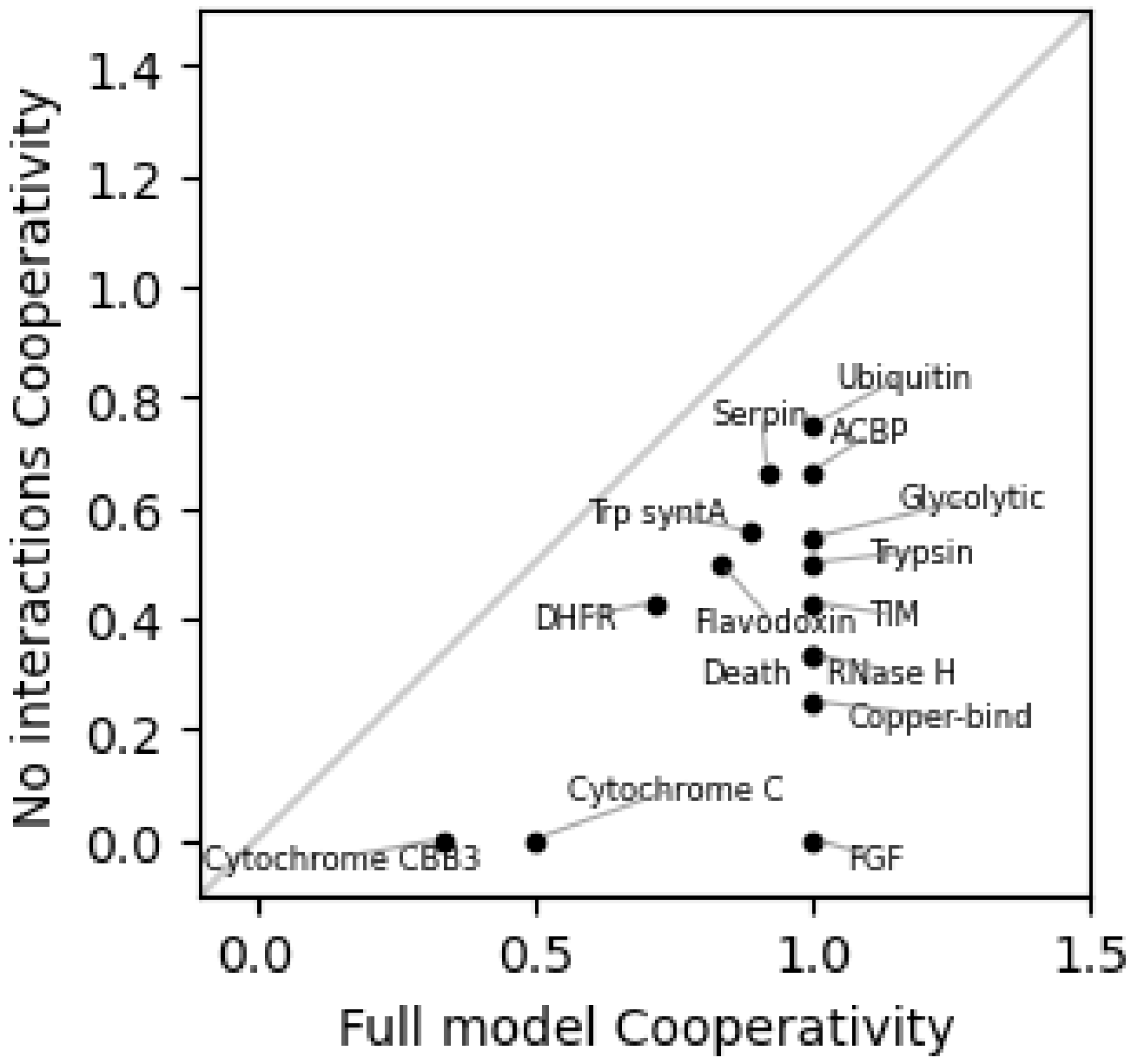

***Figure S3***. ***No-interaction control.*** *Cooperativity Score for the reference sequence of each family, obtain with the full model vs the Cooperativity score for the same sequence obtained with the model without the interaction terms between foldons* ($\epsilon^{s} = 0$). *The x = y grey curve is included as a visual reference.*

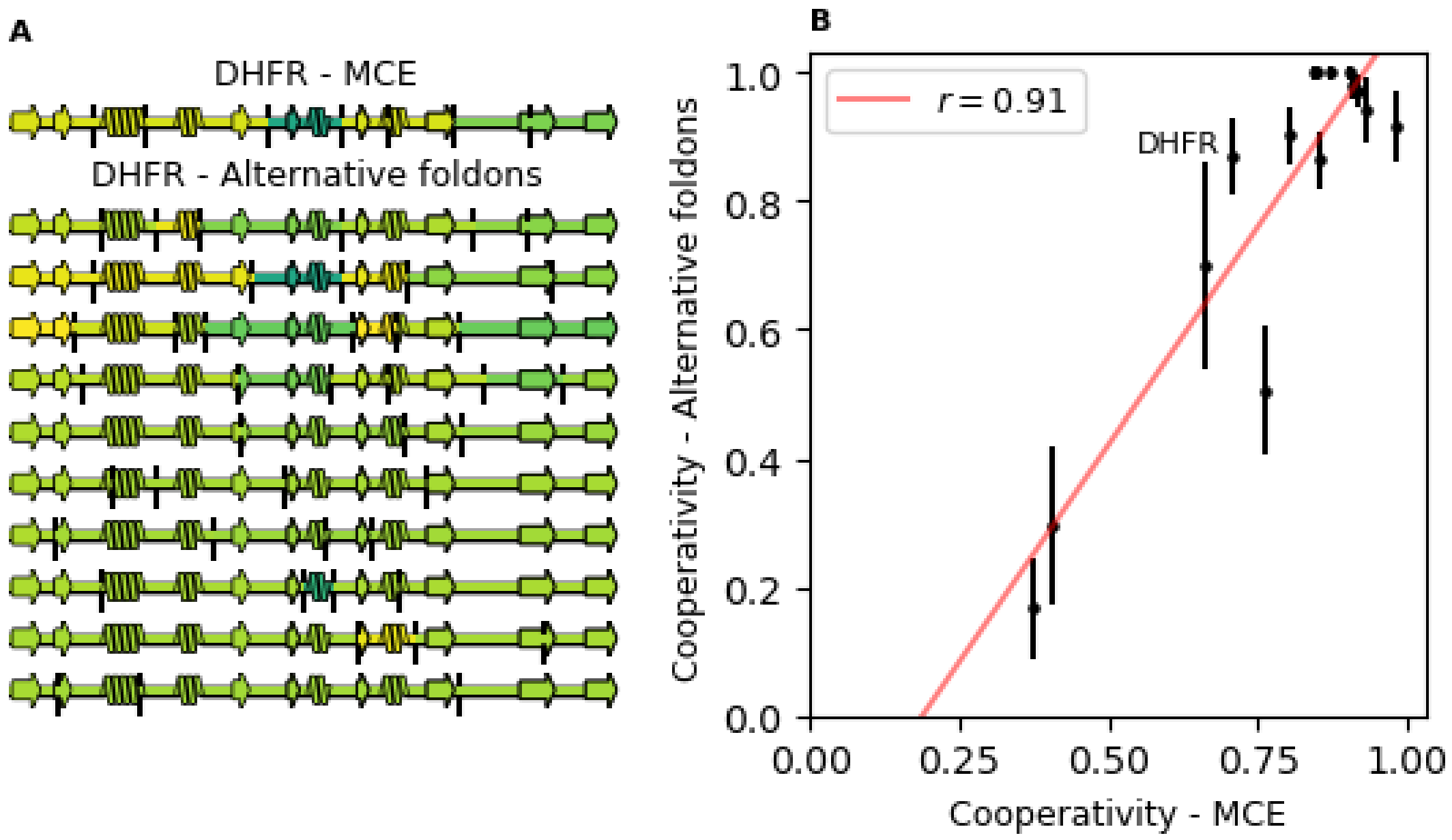

***Figure S4. Alternative foldons control.*** *(A) Secondary structure of EcDHFR colored according to element folding temperature (yellow is more stable). On top, obtained with a simulation using as foldons the Minimal Common Exons (MCE), result detailed in Fig. 2. Below, using 10 different alternative partitions. Foldons are separated by vertical dashed black lines. (B) Cooperativity obtained for the reference sequence of each family using alternative partitions vs the cooperativity for the same sequences, but using the MCE. The representative values are the averages over the alternative partitions, and the error bars the standard errors of the mean. DHFR sequence (panel A) is pointed out. The values are strongly correlated, the red line is a linear fit.*

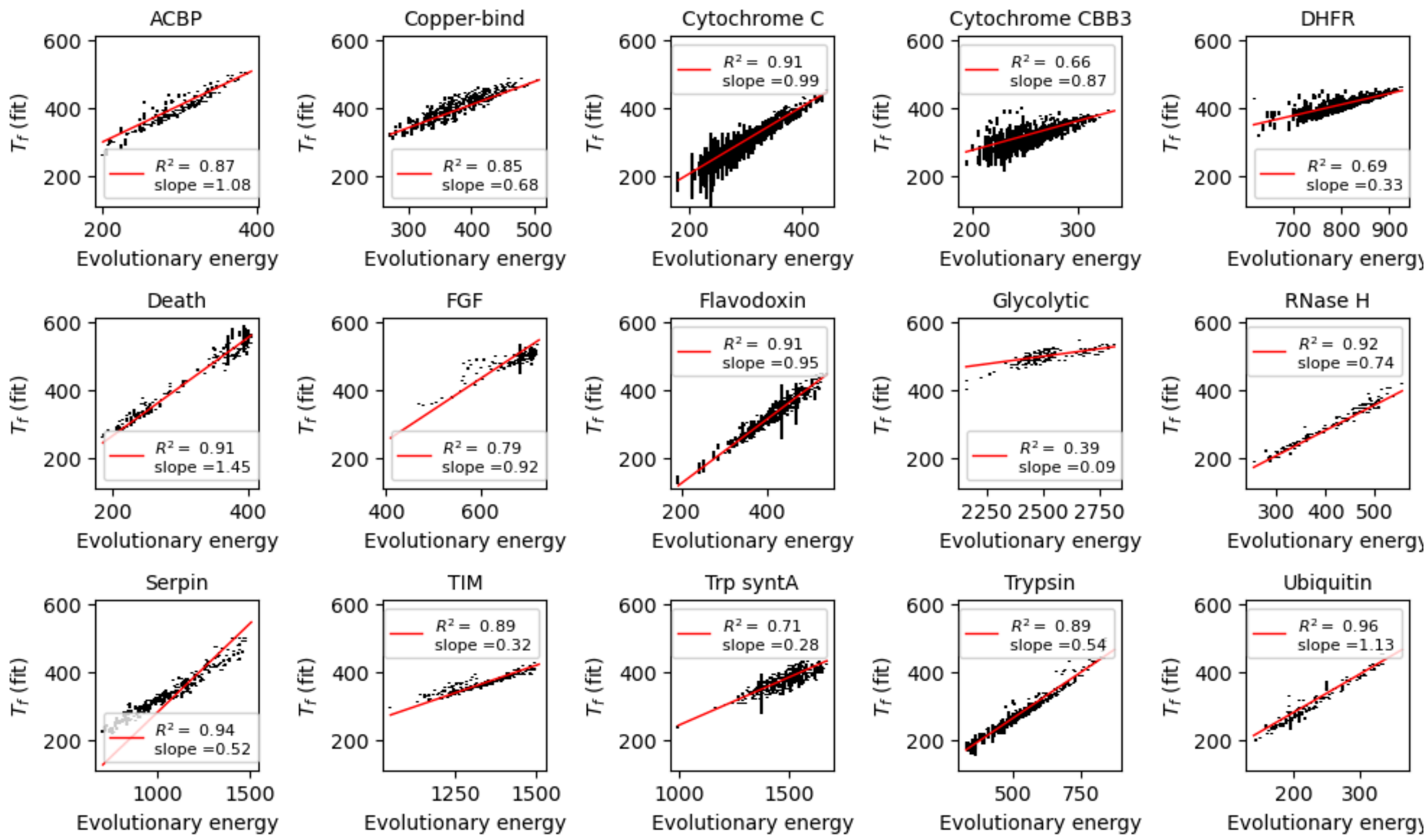

***Figure S5.*** For each family, protein Folding Temperature fitted for 500 natural sequences vs Evolutionary Energy. Linear fits are shown in red.

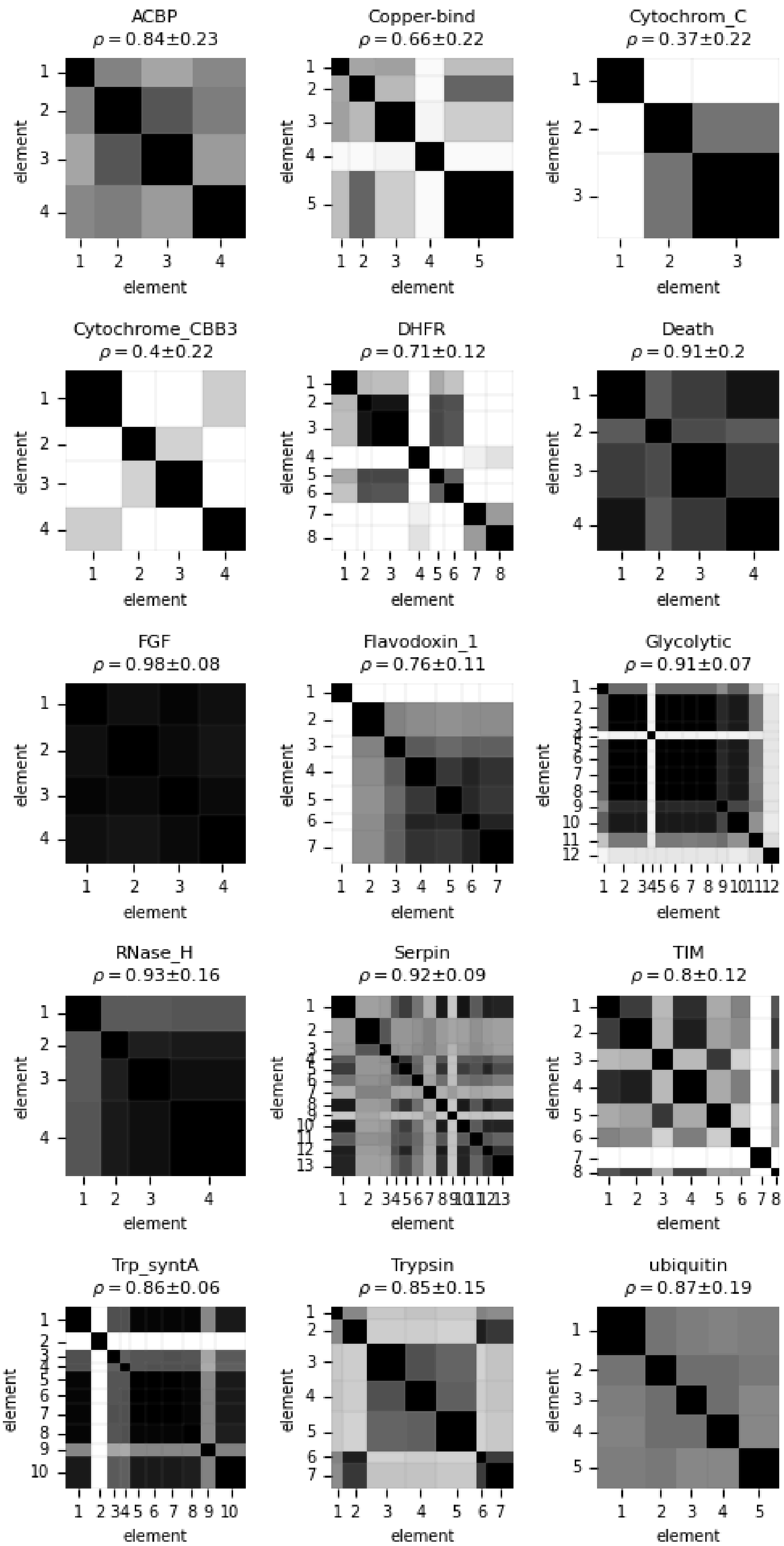

***Figure S6. Average folding apparent domains.*** *For each family, domain matrices average over 500 representative natural sequences. The color scale represents the occurrence of each subdomain within the family (black is always present, white is never present). The Cooperativity Score (ρ) average and standard deviation are indicated in each case on top of the domain matrix.*

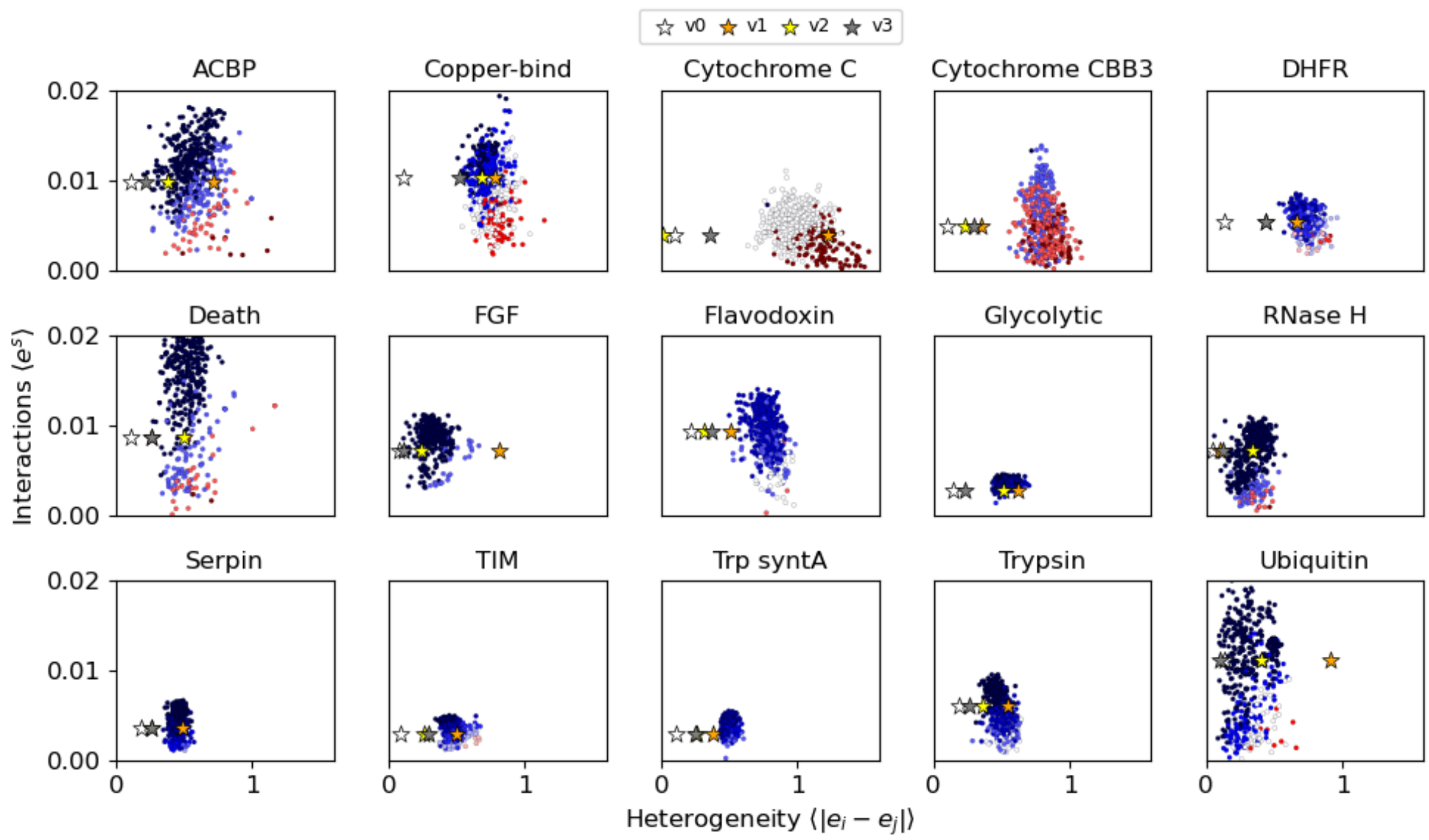

***Figure S7***. *For each family, Cooperativity score for 500 representative sequences is shown in a color scale on a plane defined by the energetic heterogeneity of foldons and their average interaction strength (see Methods). Vanilla models are marked as stars.*

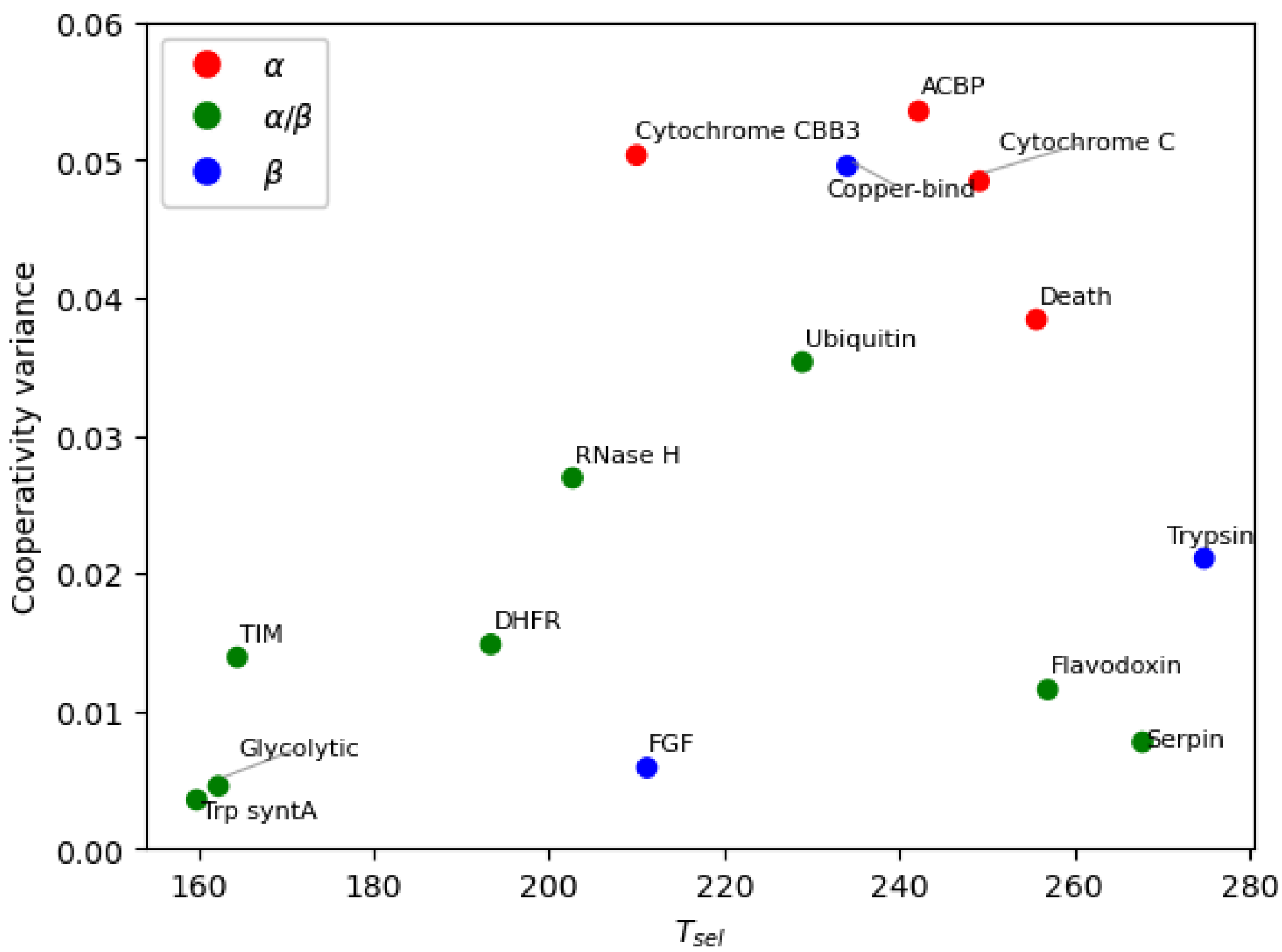

***Figure S8*** *Cooperativity variance of natural sequences as a function of the Selection Temperature.*

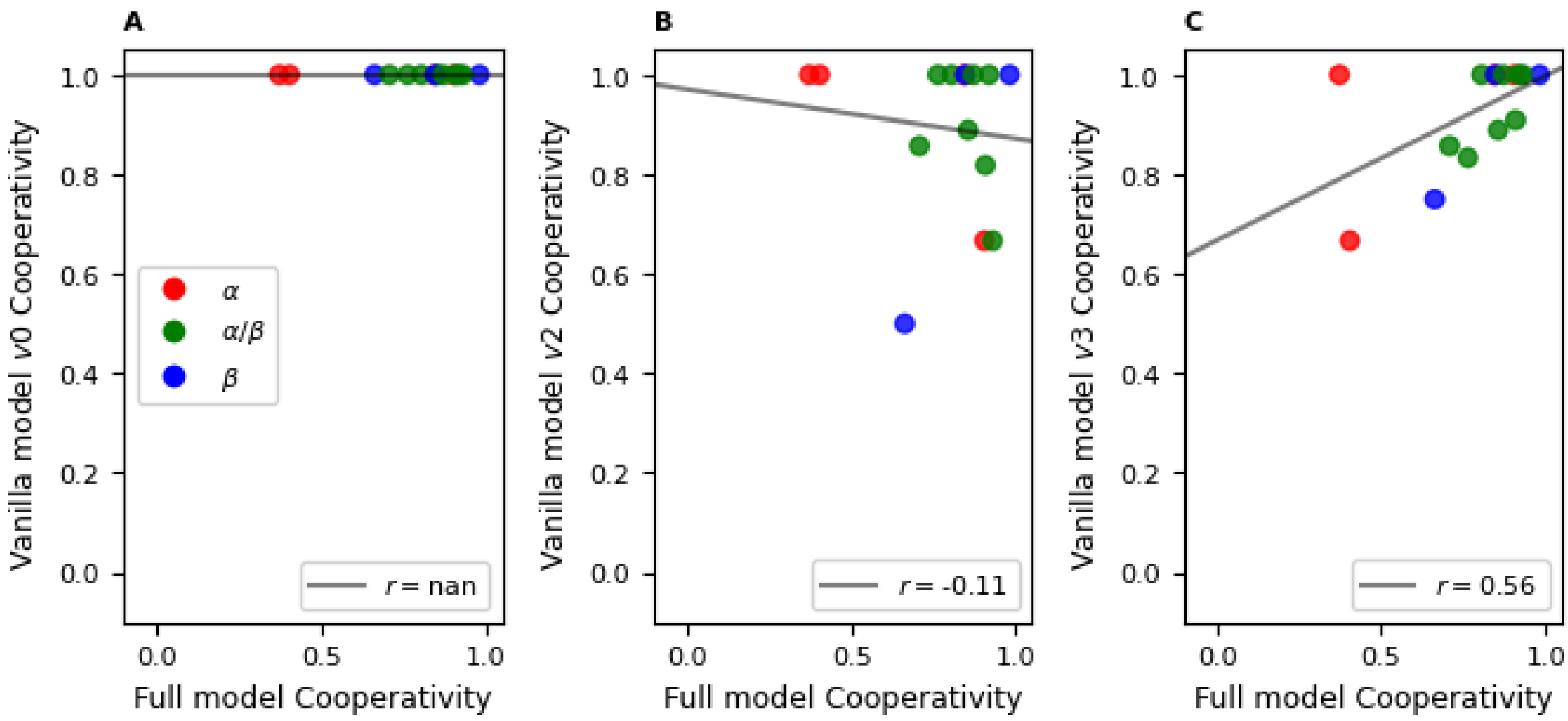

***Figure S9**. Correlation between the Cooperativity obtained with each vanilla model and, on average, with the full model for each family. The color code is the same for all panels. (A) Vanilla model v0. (B) Vanilla model v2. (C) Vanilla model v3.*

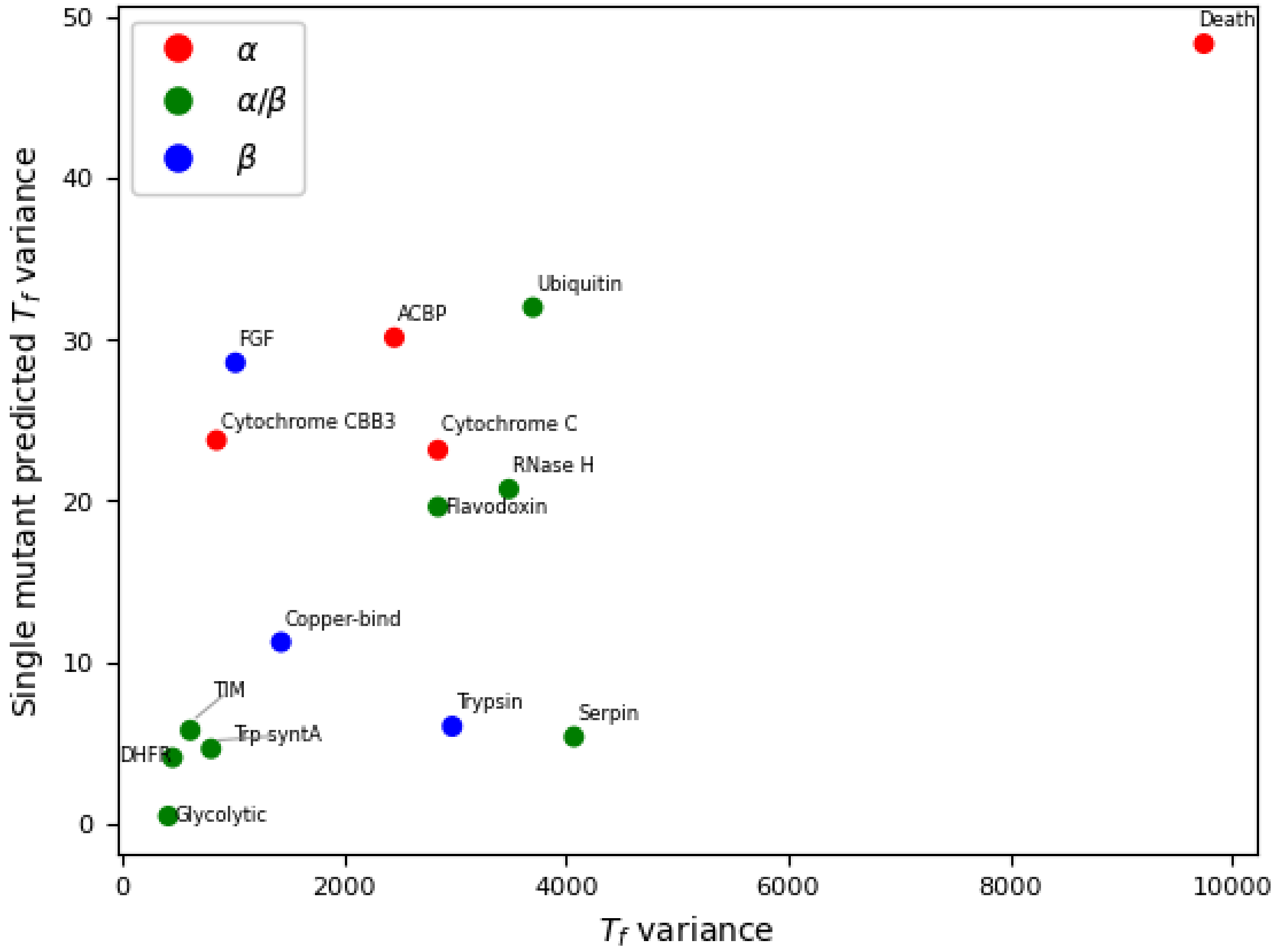

***Figure S10.*** *Variance of the predicted changes in the protein Folding Temperature ($T_f$) for point mutants as a function of the variance of $T_f$ for 500 natural sequences.*

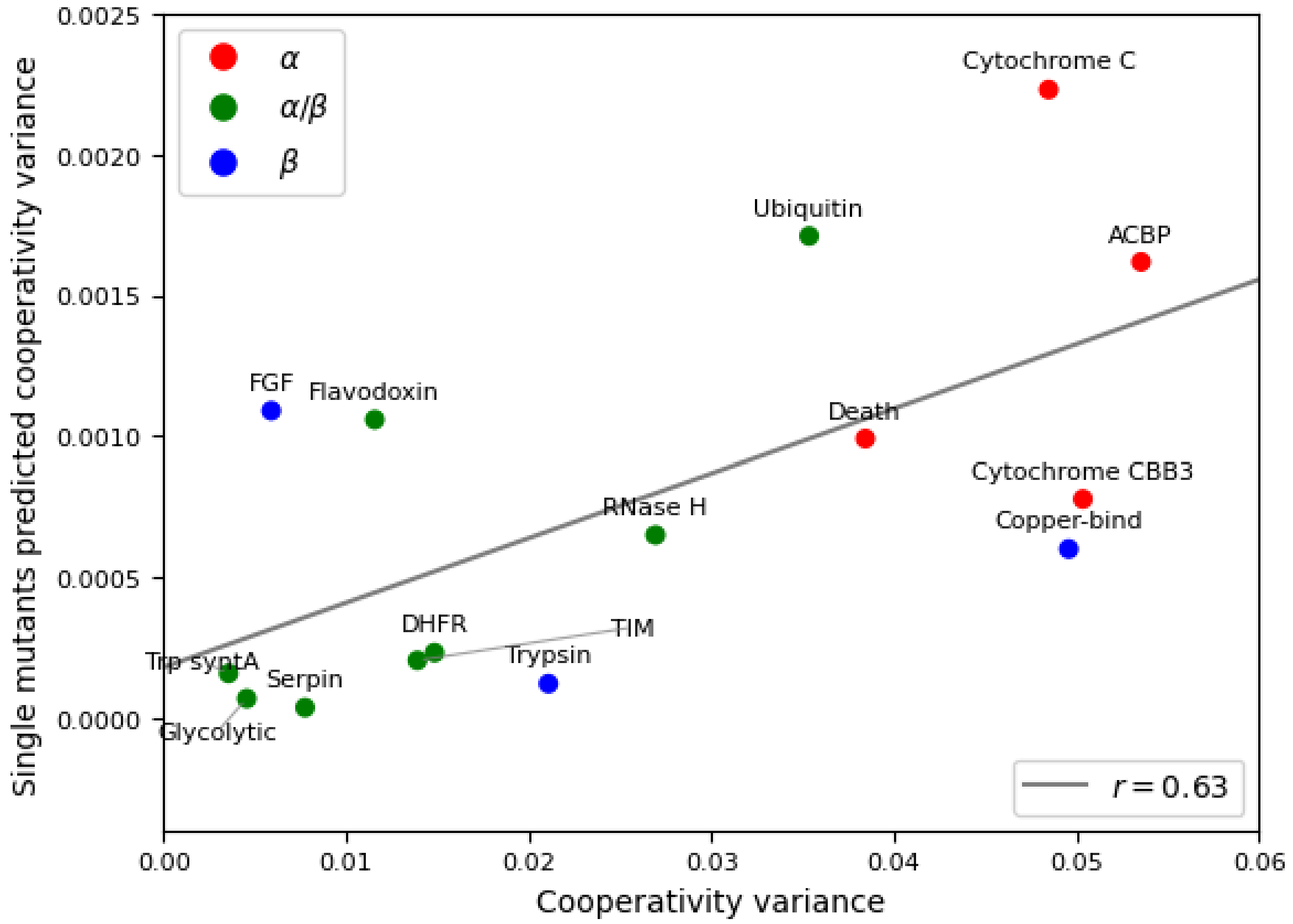

***Figure S11***. *Variance of the predicted changes in the Cooperativity for point mutants as a function of the Cooperativity variance for 500 natural sequences.*

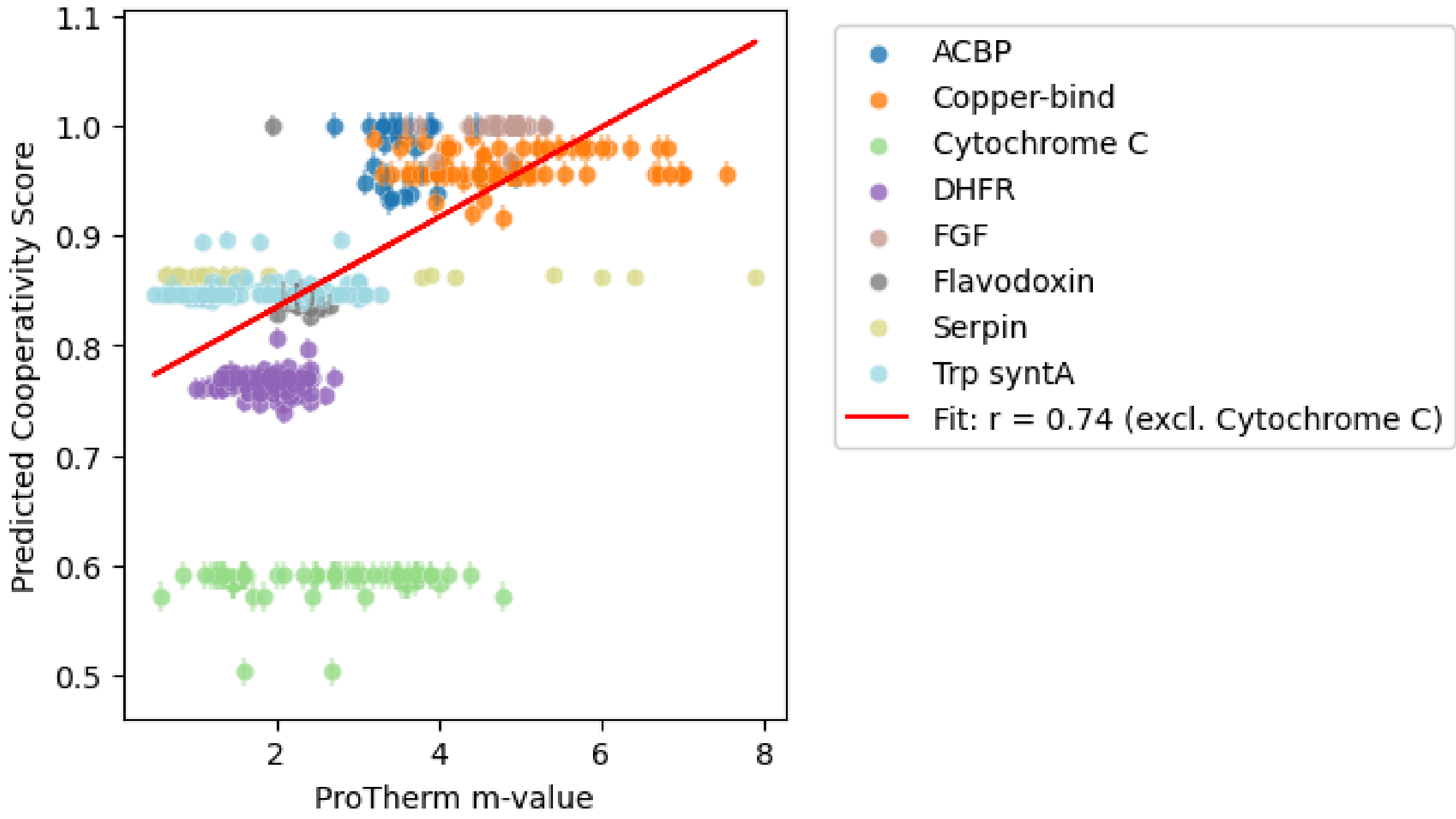

***Figure S12.*** *Correlation between the Cooperativity Score predictions and the corresponding experimental m-values from the ProTherm database.*

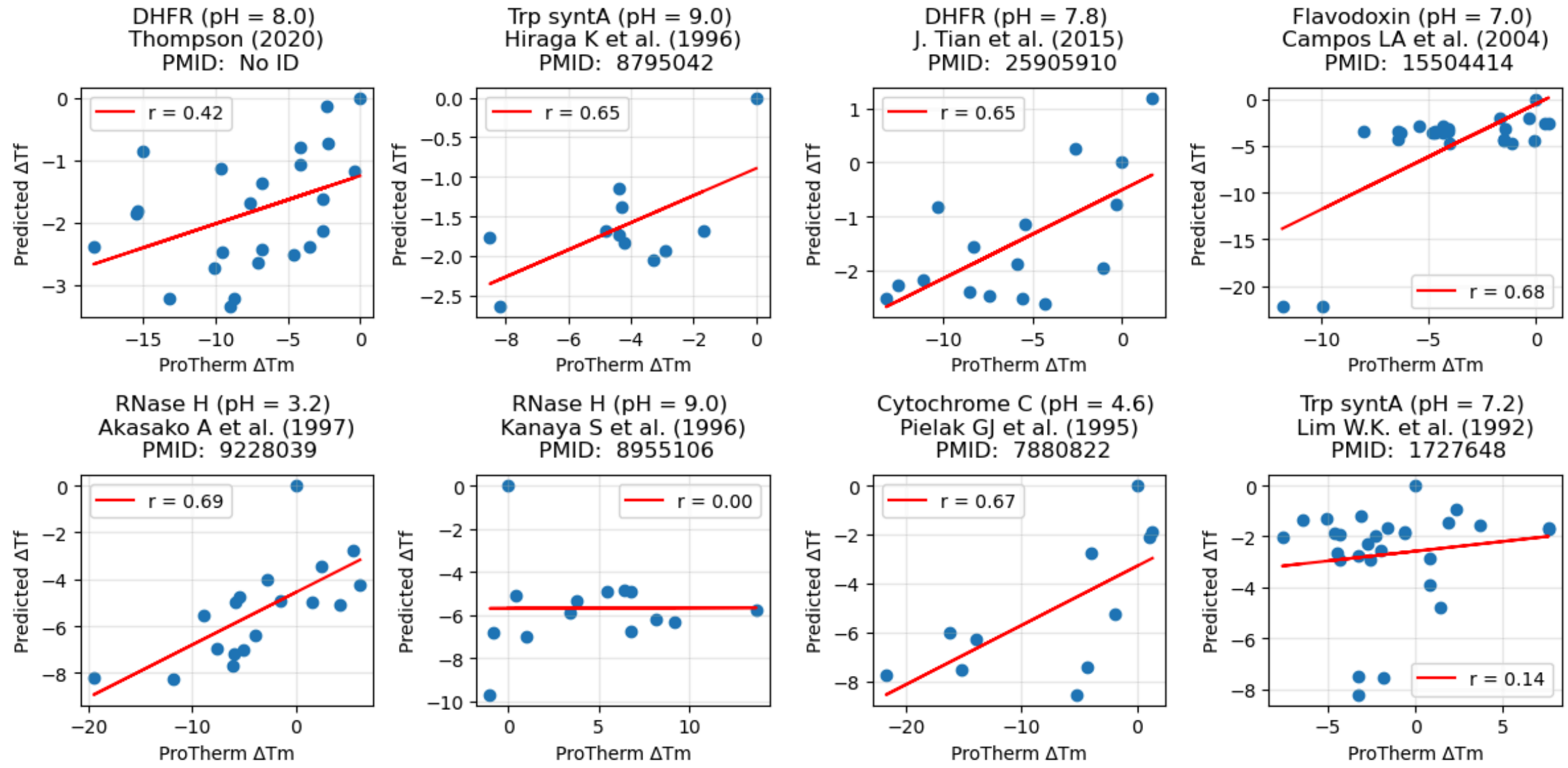

***Figure S13.*** *Correlations between the Folding Temperature change predictions and the corresponding experimental values from the ProTherm database.*

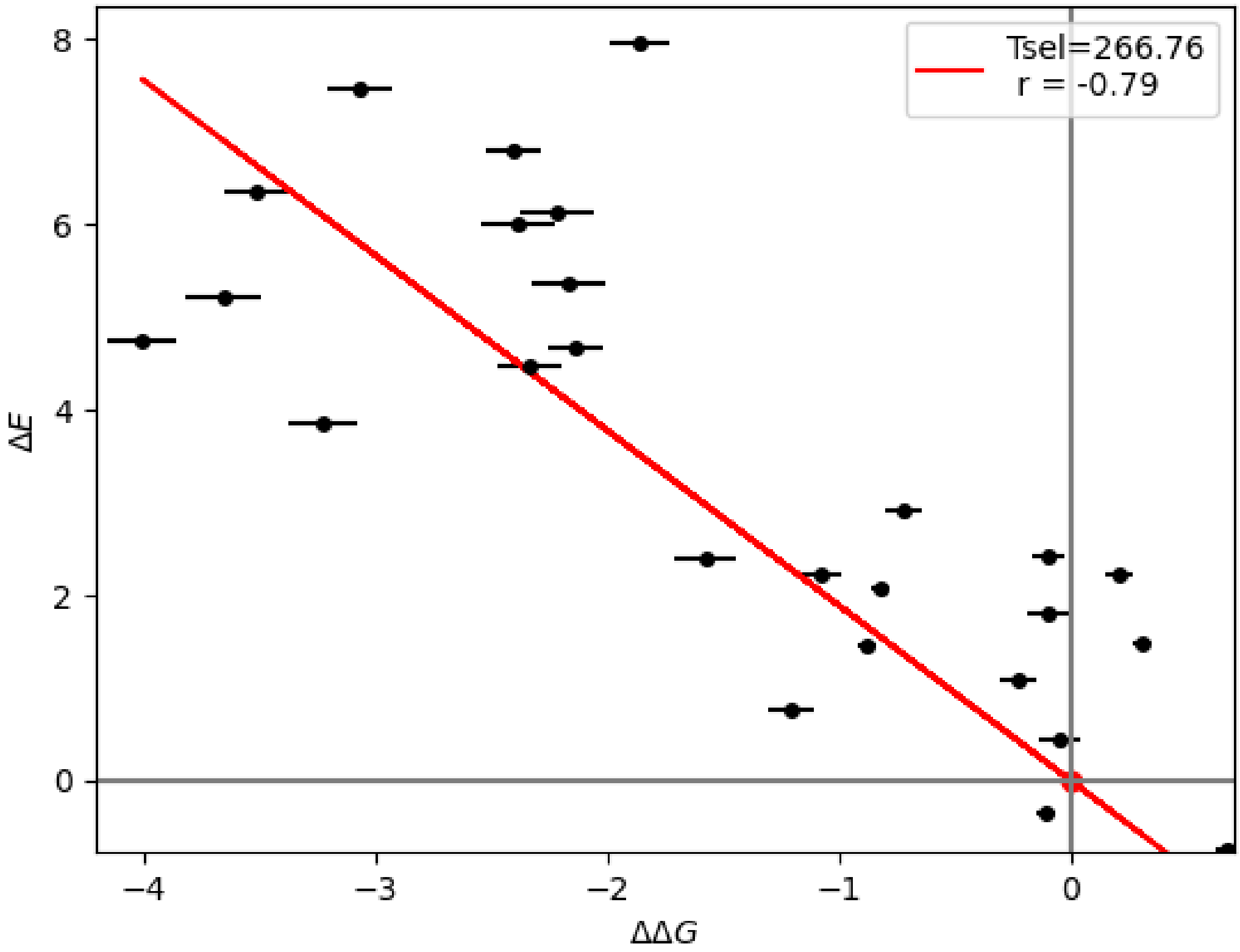

***Figure S14*. Calculation of the Selection Temperature for the reference family PDZ.**
Evolutionary Energy differences ΔE for point mutants of PDZ obtained with a RBM vs the corresponding available experimental $\Delta\Delta G$ data (see Methods). The red line is a linear fit.

**SI References**